\def\av{$A_{\rm V}$}
\def\ha{H$\alpha$}
\def\wha{$W_{\rm H\alpha+}$}
\def\dn{$D(4000)$}
\def\nt{[N\,{\sc ii}]}
\def\st{[S\,{\sc ii}]}
\def\ot{[O\,{\sc iii}]}

\documentclass[iop,appendixfloats]{emulateapj}
\usepackage{color}

\shorttitle{\ha\ and 4000~\AA\ Breaks for $\sim$3500 Distant Galaxies}
\shortauthors{Kriek et al.}

\begin{document}
  
\title{\ha\ and 4000 \AA\ Break Measurements for $\sim$3500 $K$-selected Galaxies at $0.5<\lowercase{z}<2.0$}

\slugcomment{Accepted for publication in ApJ}

\author{Mariska Kriek\altaffilmark{1}, Pieter G. van Dokkum\altaffilmark{2}, Katherine E. Whitaker\altaffilmark{2}, Ivo Labb\'e\altaffilmark{3}, Marijn Franx\altaffilmark{3}, \& Gabriel B. Brammer\altaffilmark{4}}

\altaffiltext{1}{Harvard-Smithsonian Center for Astrophysics, 60 Garden Street, Cambridge, MA 02138, USA}

\altaffiltext{2}{Department of Astronomy, Yale University, New Haven, 
  CT 06520, USA}

\altaffiltext{3}{Leiden Observatory, Leiden University, NL-2300 RA Leiden, 
  Netherlands}

\altaffiltext{4}{European Southern Observatory, Alonso de Cordova 3107,
  Casilla 19001, Vitacura, Santiago, Chile}

\begin{abstract}
 We measure spectral features of $\sim$3500 $K$-selected galaxies at
 $0.5<z<2.0$ from high quality medium-band photometry using a new
 technique. First, we divide the galaxy sample in 32 subsamples based
 on the similarities between the full spectral energy distributions
 (SEDs) of the galaxies. For each of these 32 galaxy types we
 construct a composite SED by de-redshifting and scaling the observed
 photometry. This approach increases the signal-to-noise ratio and
 sampling of galaxy SEDs and allows for model-independent stellar
 population studies. The composite SEDs are of spectroscopic quality,
 and facilitate -- for the first time -- H$\alpha$ measurement for a
 large magnitude-limited sample of distant galaxies. The linewidths
 indicate a photometric redshift uncertainty of
 $\Delta z<0.02\times(1+z)$. The composite SEDs also show the Balmer and
 4000\,\AA\ breaks, Mg\,{\sc ii} absorption at $\sim$2800~\AA, the
 dust absorption feature at 2175\,\AA, and blended \ot+H$\beta$
 emission. We compare the total equivalent width of \ha, \nt, and \st\
 (\wha) with the strength of the 4000~\AA\ break (\dn) and the
 best-fit specific star formation rate, and find that all these
 properties are strongly correlated. This is a reassuring result, as
 currently most distant stellar population studies are based on just
 continuum emission. Furthermore, the relation between \wha\ and \dn\
 provides interesting clues to the SFHs of galaxies, as these features
 are sensitive to different stellar ages. We find that the correlation
 between \wha\ and \dn\ at $0.5<z<2.0$ is similar to $z\sim0$, and
 that the suppression of star formation in galaxies at $z<2$ is
 generally not abrupt, but a gradual process. 
\end{abstract} 


\section{INTRODUCTION}\label{sec:int}

Galaxies in the local universe give us many clues to their star
formation histories (SFHs). Massive elliptical galaxies can have
stellar ages nearly as old as the age of the universe, while spiral
galaxies are still forming new stars. Detailed studies of nearby
stellar populations allow accurate age measurements, and thus a
measure of the SFHs in these systems. While such detailed studies will
never be possible at cosmological distances, studying galaxies
throughout cosmic time offers another advantage: we can directly
study their masses and star formation rates (SFRs) at different phases
in their lives, rather than inferring these quantities from age
measurements.

In the past few years, photometric and spectroscopic galaxy surveys
have started to explore the universe at redshifts $z=1-3$, the peak of
the quasar era and the epoch when the volume-averaged SFR of the
Universe was at its maximum \cite[e.g,][]{bo10}. This peak is at least
in part caused by a population of galaxies with apparently very high
specific SFR (SSFR; SFR divided by stellar
mass) \citep[e.g.,][]{no07a,no07b,da07,sa07,el07,da09,ol10,pa11}. Remarkably,
at the same time there are galaxies that appear to have already
stopped forming stars. Massive, quiescent galaxies have been
identified out to at least
$z\sim2.5$ \citep[e.g.,][]{mc04,la05,kr06b,kr08b,ci08,wi09,br09,wh10},
implying that they formed their stars at even earlier times.

A major complication in all high-redshift studies is the lack of
uniform and reliable measurements of SFHs. In the local universe, the
``standard'' indicator of the instantaneous SFR is the H$\alpha$
emission line \citep[e.g.,][]{ke98} and a standard indicator of the age of a
stellar population is the strength of the  4000\,\AA\ HK continuum
break \citep[e.g.,][]{ka03}. At redshifts $z>1$, both features shift
into the observed near-infrared (NIR) and are very difficult to
measure. As a result, studies of galaxies at $z>1$ are either limited
to small and often biased samples
\citep[e.g.,][]{er06a,er06b,er06c,kr06a,kr07,kr08a,mu09,mu10,re10}
or are based on SFH indicators that are less robust and less well
calibrated, such as broadband spectral energy distributions
(SEDs) \citep[e.g.,][]{vd06,la07,pa07,pa11,da09}.

As a compromise between sample size and higher spectral resolution, we
have undertaken the NEWFIRM medium-band
survey \citep[NMBS;][]{vd09,wh11} in the COSMOS \citep{sc07} and
AEGIS \citep{da07} fields. The ``high'' resolution NIR photometry in
combination with the public data available in these fields provides
accurate photometric redshifts and stellar population properties out
to $z\sim3$ \citep[][]{br09,vd10,wh10,kr10,ma10,wa11,br11}. In the
present work we identify groups of analogous galaxies at $0.5<z<2.0$
and construct composite SEDs. This approach increases the
signal-to-noise and sampling of galaxy SEDs and allows for
model-independent stellar population studies. The SEDs are of
exquisite quality, showing spectral features, such as the \ha+\nt+\st\
and H$\beta$+[O\,{\sc iii}] emission lines, the Balmer or 4000\,\AA\
breaks, Mg{\sc\,ii} absorption at 2800~\AA, the continuum break at
2640~\AA, and the dust absorption feature at 2175\,\AA. Thus, these
SEDs can be used to study the stellar populations and emission line
characteristics of typical galaxies in a very detailed way, which is
not possible when the photometry of individual galaxies is considered
separately.

This first paper describes our technique, provides an overview of all
composite SEDs and example rest-frame optical morphologies, and
studies the correlations between the equivalent width (EW) of \ha, the
strength of the 4000~\AA\ break, and the SSFRs of $\sim 3500$ galaxies
at $0.5<z<2.0$. Additionally, we compare our measurements to those of
low-redshift galaxies from the SDSS and stellar population synthesis
(SPS) models, and discuss the implications for the SFHs of
galaxies. Other applications of the composite spectra, including the
relation between galaxy size and SED type, dust properties, and the
demographics of active galactic nuclei (AGNs) will be discussed in
future papers.

\section{CONSTRUCTING COMPOSITE SEDS}\label{construct}

The galaxy sample used to construct the composite SEDs is selected
from the NMBS. The NMBS uses five custom NIR filters and covers a
total area of 0.4 square degrees in the COSMOS \citep{sc07} and
AEGIS \citep{da07} fields. The medium-band filter set is combined with
publicly available imaging at FUV-to-MIR
wavelengths \citep{er09,hi09,ba08,san07,ca07} as described
in \cite{wh11}. For the work presented in this paper, we limit our
sample to the COSMOS field, as medium-band optical photometry from
Subaru is available for this field as well \citep[e.g.,][ Y. Taniguchi
et al. in preparation]{ta07,il09}. The high photometric sampling
provides accurate photometric redshifts and stellar population
properties \citep[e.g.,][]{br09,vd10,wh10}. The photometric redshifts
and stellar population properties are derived using EAZY \citep{br08}
and FAST \citep{kr09a}, respectively. In order to limit our sample to
high-quality SEDs with rest-frame UV-to-NIR wavelength coverage, we
require a redshift of $0.5<z<2.0$ and a signal-to-noise (S/N) of 25 in
the $K$-band.

\begin{figure}                     
  \begin{center} 
  \includegraphics[width=0.45\textwidth]{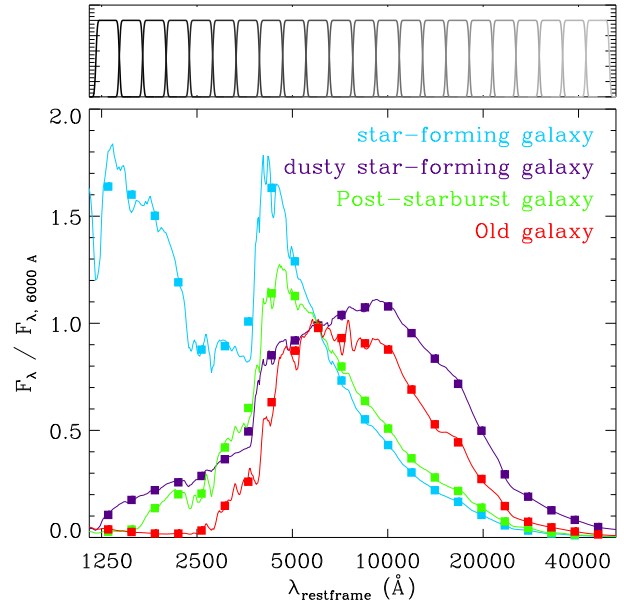}\hspace{0.2in}

    \caption{{\it Top:} The 22 synthetic rest-frame filters used to
      identify galaxies with similar SED shapes. {\it Bottom:}  SEDs
      and 22 rest-frame colors for four different  galaxy types. This
      panel illustrates that using these 22 bands we can discriminate
      between different types and find galaxies with similar SED
      shapes and thus properties. For example, the value for $b$
      between the two closest SEDs in this figure (the dusty
      star-forming in purple and old galaxy in red) is
      0.23. \label{filters}}

  \end{center} 
\end{figure}

We define 22 artificial rest-frame filters of intermediate bandwidth,
evenly spaced in log$\,\lambda$ between 1200\,\AA\ and 50\,000\,\AA,
and compute the fluxes in these filters for all galaxies in our
sample, using the method described by \cite{br09}. In
Figure~\ref{filters} we show the filters in comparison to the SEDs of
different galaxy types. For each galaxy in our sample we identify
analogous galaxies, using the following expression:
\begin{equation}\label{eq:corr}
\sqrt{\frac{\Sigma~(f_{\lambda}^{\rm ob1} - a_{\rm 12} f_{\lambda}^{\rm ob2})^2}{\Sigma~(f_\lambda^{\rm ob1})^2}} ~~ = ~~ b_{12}
\end{equation}
with $a$ the scaling factor
\begin{equation}\label{eq:scale}
a_{\rm 12} = \frac{\Sigma~f_{\lambda}^{\rm ob1} f_{\lambda}^{\rm
ob2}}{\Sigma~(f_{\lambda}^{\rm ob2})^2}
\end{equation}
and $b$ the variation between two galaxies. In practice, we generally
use fewer than 22 rest-frame filters to relate galaxies, as we only
compare the rest-frame wavelength coverage that two galaxies have in
common.

We start by taking $b<0.05$ and identify the galaxy with most analogs;
this is the primary galaxy for the first subsample. We choose this
value of $b$ as it yields distinct composite SEDs, while keeping the
intrinsic scatter per SED type low. By decreasing the value for $b$ we
found identical best-fit stellar population models for similar SED
types, while for larger values of $b$ the SEDs of individual
``analog'' galaxies were not well matched.

Next, we remove the primary galaxy and all its analogs from the parent
sample and again find the galaxy with the largest number of
analogs. We repeat this procedure until the primary galaxy has fewer
than 19 analogs. In total we find 32 subsamples with at least 20
galaxies, which include 83\% of the total galaxy sample. Some analog
galaxies that have been removed in one of the first few iterations
might have been a better match with another galaxy type which has been
constructed during a later iteration. Thus, after having divided the
galaxies into subsamples, we reassign 15\% of the galaxies which are a
better match with a primary galaxy of another subsample. For example,
several galaxies in the subsample of type 3 move to type 2 or 4 (see
Section~\ref{analysis}).
  
\begin{figure} 
  \begin{center} 
  \includegraphics[width=0.46\textwidth]{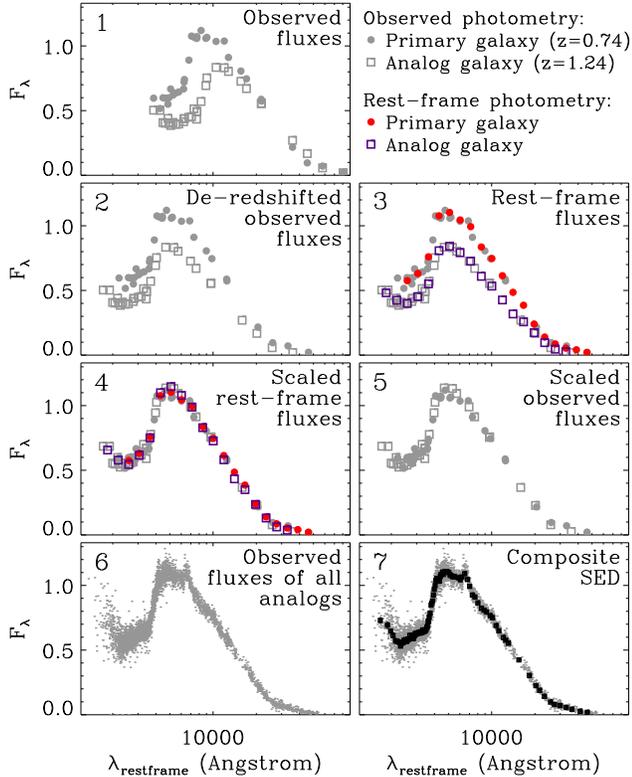}\hspace{0.2in}

    \caption{Illustration of our method to construct composite
      SEDs. In panel 1 we show the primary galaxy and one of its
      analogs. We de-redshift both galaxies to the rest-frame (panel
      2) and determine the fluxes for the artificial filters, as shown
      in Figure 1 (panel 3). Next, we scale the analog spectrum to its
      primary using the fluxes of the rest-frame filters they have in
      common (panel 4). In panel 5 we show the combined scaled and
      de-redshifted observed photometry of the primary and analog
      galaxy. We do the same for all analogs, and obtain the combined
      photometry in panel 6. Finally, we construct a composite SED by
      averaging the scaled, de-redshifted observed fluxes in
      wavelength bins.\label{ill}}

  \end{center} 
\end{figure}

\begin{figure} 
  \begin{center} 
    \includegraphics[width=0.46\textwidth]{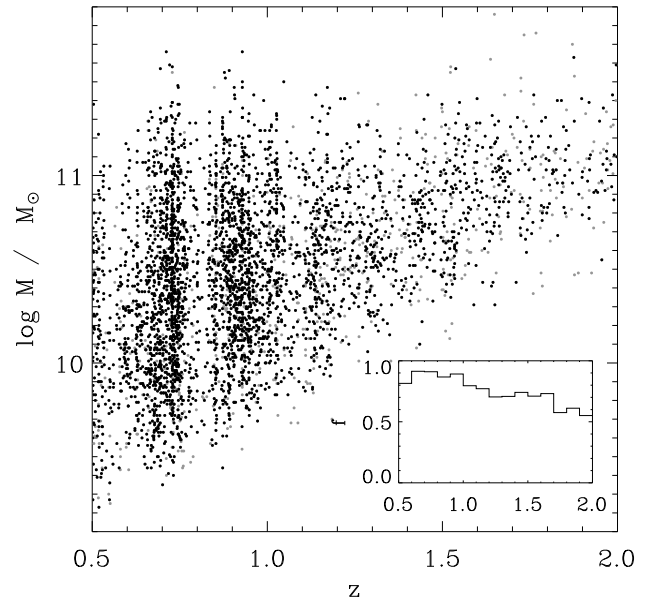}\hspace{0.2in}

    \caption{Stellar mass vs. redshift for all 4235 galaxies in the
      NMBS-COSMOS field with $0.5<z<2.0$ and a S/N of $>25$ in the
      $K$-band. The 83\% of the galaxies that are included in the 32
      SED types are shown in black, while the remaining 17\% are
      indicated by the gray symbols. In the inset we show the fraction
      of galaxies that is included as a function of redshift. In
      Appendix A we further discuss the completeness of our SED
      sample.\label{mass}}

  \end{center} 
\end{figure}

\begin{figure*}
  \begin{center} 
    \includegraphics[width=0.98\textwidth]{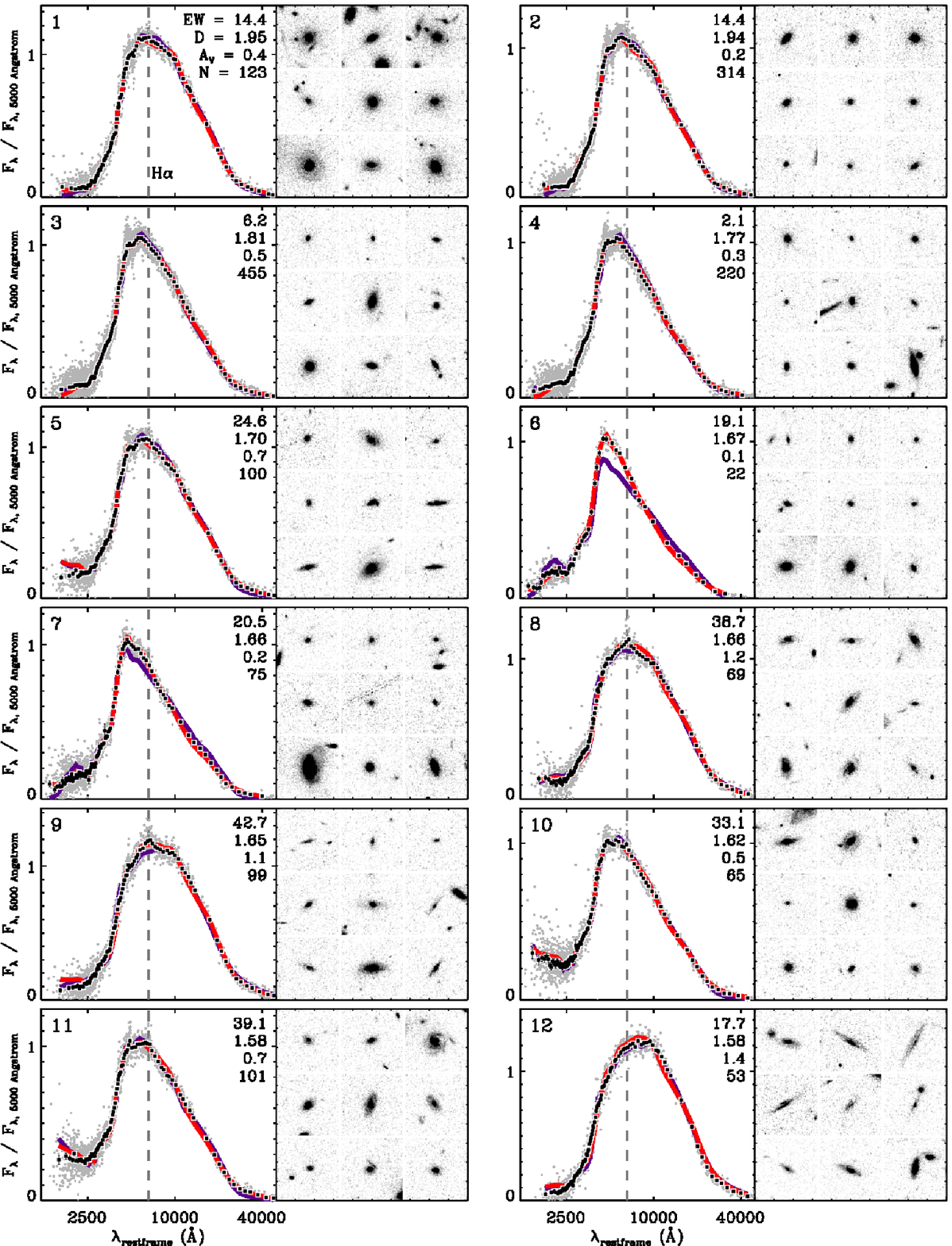} 
  \end{center}            
\end{figure*}          

\begin{figure*}
  \begin{center} 
    \includegraphics[width=0.98\textwidth]{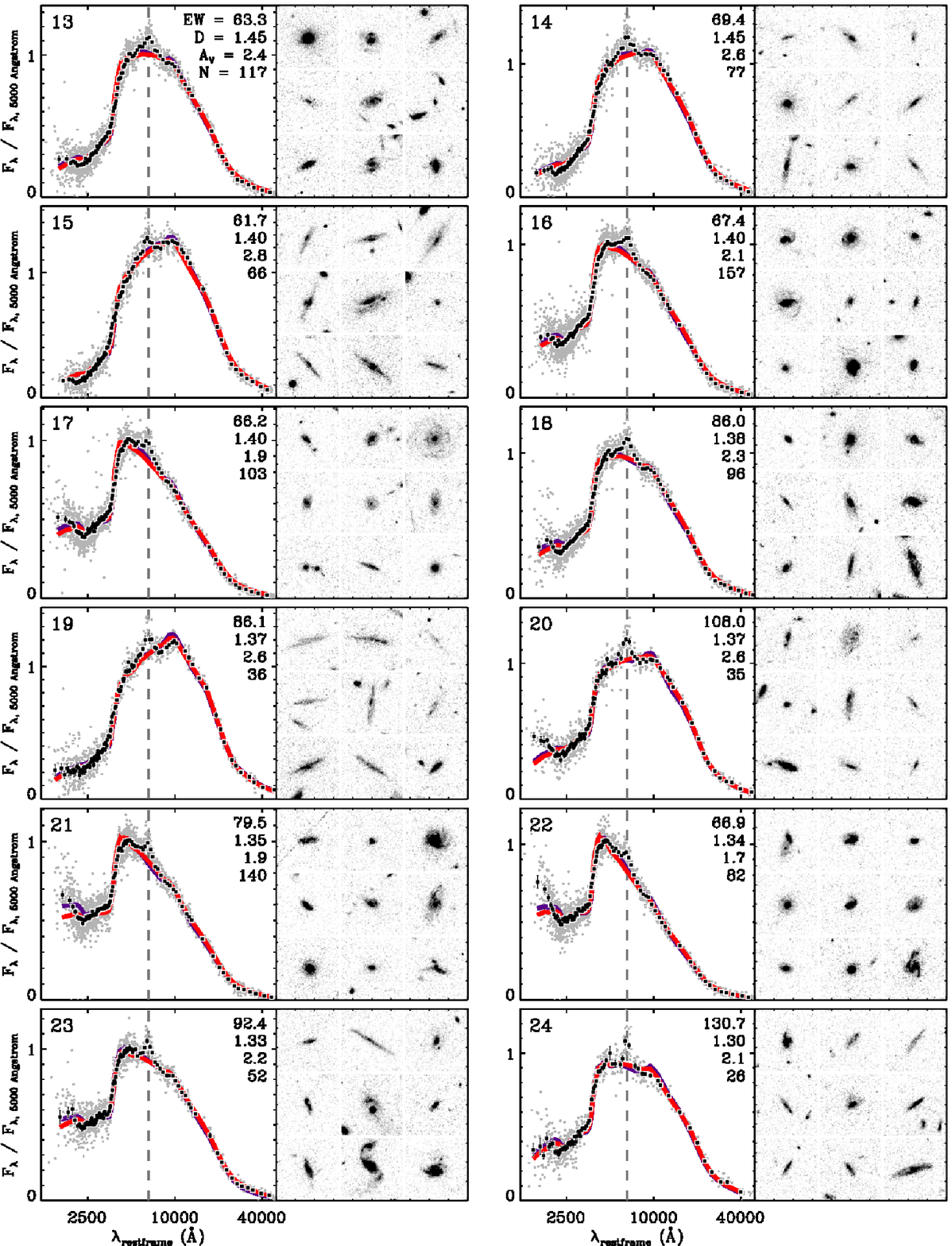} 
  \end{center}             
\end{figure*}

\begin{figure*}[!t] 
  \begin{center} 
    \includegraphics[width=0.98\textwidth]{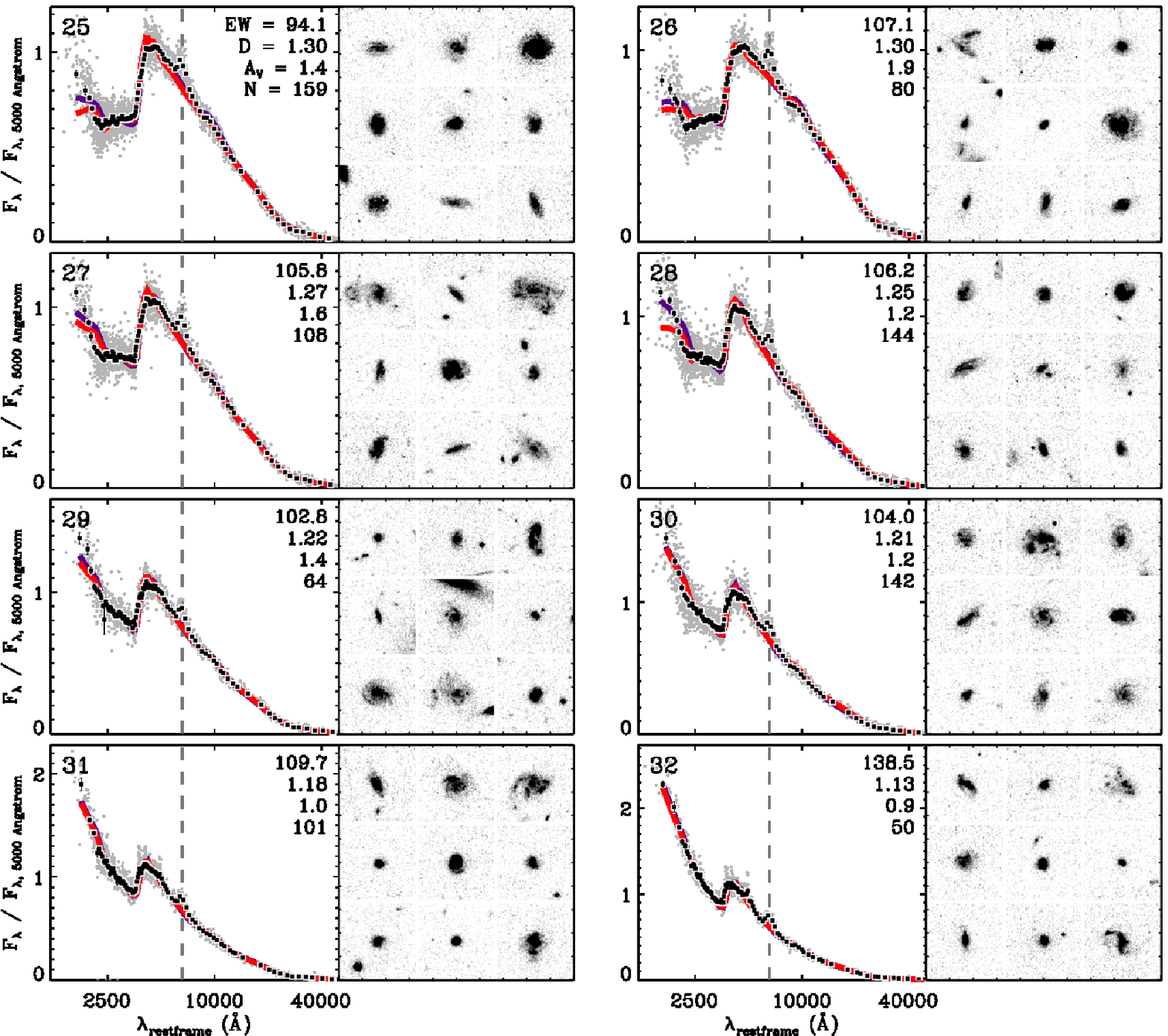} 
    \caption{Left panels: Composite SEDs constructed from the NMBS
      photometry for galaxies at $0.5<z<2.0$ with S/N$_{K} > 25$. The
      SEDs are ordered according to their \dn, starting with the SED
      type with the strongest 4000\,\AA\ break. The dashed line
      indicates the location of \ha. The red and purple curves are the
      best-fit \cite{bc03} and \cite{ma05} models, respectively. The
      value for \dn\ and \wha, the best-fit value for \av\ for
      the \cite{bc03} models, and the number of galaxy included in the
      composite SED are printed in each panel. Right panels: For each
      composite SED we show the $i$-band HST/ACS images of the nine
      galaxies closest in redshift to $z=0.9$, thus probing
      $\lambda\sim$ 4000\,\AA\ in rest-frame. \label{compilation}} 
  \end{center}
\end{figure*}                 

We create a composite SED for each subsample, by de-redshifting all
galaxies, and scaling the observed photometry of the analogs to that
of the primary galaxy. Our method is illustrated in
Figure~\ref{ill}. Note, that we use {\it observed} photometry in the
stacks, not the 22 rest-frame filters; the 22 filters are only used to
find analogs. We construct the composite SEDs by averaging the
rest-frame scaled fluxes of the primary and analog galaxies in
wavelength bins. The number of data-points per wavelength bin depends
on the number of analog galaxies. For each bin we calculate both the
error on the mean (using bootstrap resampling) and the scatter in the
distribution. To determine the scatter we first derive and remove the
running median from all combined photometric data points. Next, we
derive the scatter within each bin using the residuals. Finally, we
determine the effective filter curve for each bin by adding the
de-redshifted, normalized (by total integrated transmission) filter
curves of all included data points.

As mentioned above, the composite SEDs include 83\% of the total
galaxy sample. In Figure~\ref{mass} we show the parent sample and
indicate the 83\% that we use in the composite SEDs as a function of
stellar mass and redshift. The inset shows the completeness fraction
as a function of redshift. The figure clearly illustrates that our
sample is weighted towards lower redshift, with a median redshift of
$z=0.89$.  In Appendix A, we further discuss the completeness of the
sample and what types of galaxies we may be missing.

\section{ANALYSIS}\label{analysis}

\begin{figure}  
  \begin{center}           
    \includegraphics[width=0.48\textwidth]{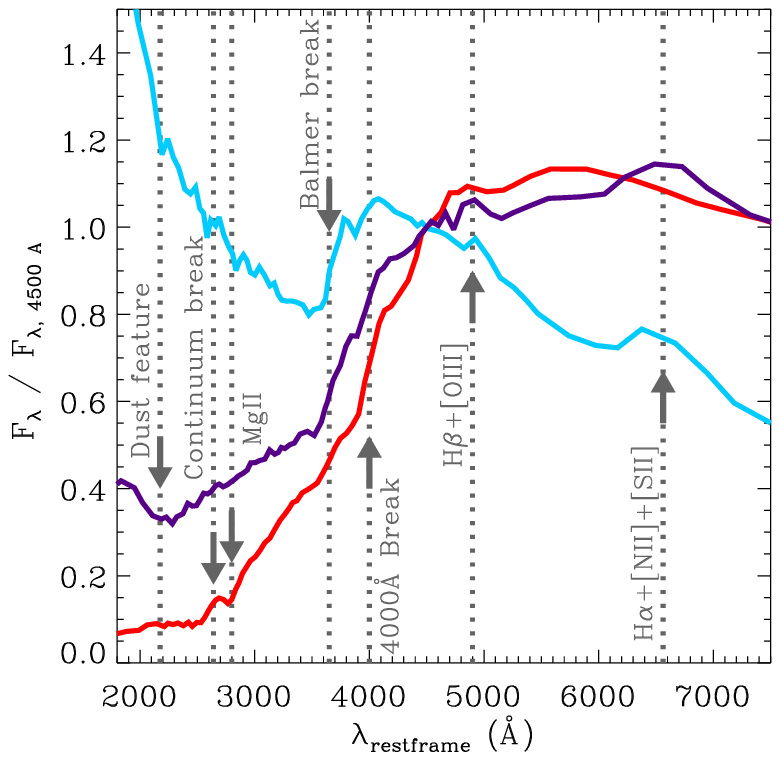} 

    \caption{Zoom on the spectroscopic features of three different SED
      types. For the quiescent type (red) we detect the 4000\,\AA\
      break, Mg{\sc\,ii} absorption at 2800~\AA, and the continuum
      break at 2640~\AA. For the dusty star-forming type (purple) we
      detect the dust absorption feature at 2175\,\AA,
      the \ha+\nt+\st\ emission line and possibly the
      H$\beta$+[O\,{\sc iii}] emission line. For the less obscured
      star-forming type (blue) we detect the Balmer break, and the two
      sets of blended emission lines. \label{features}}

  \end{center} 
\end{figure}

The composite SEDs are presented in Figure~\ref{compilation}.  They
are of exquisite quality, showing spectral features, such as
the \ha+\nt+\st\ and H$\beta$+[O\,{\sc iii}] emission lines, the
Balmer or 4000\,\AA\ breaks, Mg{\sc\,ii} absorption at 2800~\AA, the
continuum break at 2640~\AA, and the dust absorption feature around
2175\,\AA. In Figure~\ref{features} we indicate all these feature in 3
different SED types: a blue star-forming galaxy type, a dusty
star-forming galaxy type, and a quiescent galaxy type. In this section
we will study the stellar populations of the different SED types using
spectral indices and by comparing the full SED shapes with SPS models.

\subsection{SPS fitting}

In this section we derive stellar population properties by comparing
the composite SEDs with SPS models. We fit the composite SEDs by both
the \cite{bc03} and \cite{ma05} SPS models, assuming a delayed
exponential SFH of the form $\psi(t)\propto t \,{\rm exp}(-t/\tau)$,
and leaving age ($t$), the e-folding time ($\tau$), the amount of dust
attenuation (\av), and metallicity ($Z$) as free parameters (see note
to Table~\ref{tab:mod}). We assume the \cite{ca00} attenuation curve,
which we implement as a uniform screen. We adopt a \cite{sa55} initial
mass function (IMF). Compared to a \cite{kr01} or \cite{ch03} IMF our
choice will primarily affect the mass-to-light ratio, and has little
impact on other stellar population properties. We mask the bins that
are possibly contaminated by \ha\ while fitting. For the photometric
uncertainties we take the scatter, and thus our confidence intervals
on the stellar population properties essentially reflect the variation
within a galaxy subsample.

The best fits for the \cite{bc03} and \cite{ma05} SPS models are
represented by the red and purple curves in Figure~\ref{compilation},
respectively. The corresponding best-fit values and confidence
intervals are given in Table~1. The reduced $\chi^2$ values indicate
that most SED types are equally well fit by both models, though the
stellar population properties implied by the fits are not always
consistent with each other (see \cite{co09,co10,cg10} for a detailed
discussion on the uncertainties in SPS models). The exception though
are post-starburst galaxies, which are represented by SED type 6 (and
to a lesser extent type 7). As shown in \cite{kr10}, the \cite{ma05}
models overestimate the contribution of the thermally-pulsing
asymptotic giant branch (TP-AGB) stars at NIR wavelengths \citep[see
also][]{cg10}. SED type 12 is poorly described by both models, though
it is not clear why. These galaxies might have been placed at the
wrong redshift, resulting in a SED shape which cannot be explained by
any SPS model. This will be explored in detail in I. Labb\'e et
al. (in preparation).

Many SEDs are poorly fit in the rest-frame UV. This is likely due to
the presence of the dust bump at 2175~\AA\ \citep{st65}, which is not
included in the used attenuation model. Although this feature is seen
in the Milky Way and several other galaxies (even at high redshift),
the exact origin is still unknown \citep[e.g.,][]{dr93,co10b}. We will
further investigate this in a future paper.

\subsection{\dn}

The composite SEDs allow a direct measurement of the 4000~\AA\
break. The 4000~\AA\ break is due to absorption in the atmospheres of
stars and arises because of an accumulation of absorption lines of
primarily ionized metals, among which Ca\,{\sc ii} H and K. As the
opacity increases with decreasing stellar temperature, the 4000 \AA\
break gets larger with older ages, and is largest for old and
metal-rich stellar populations. The 4000~\AA\ break is generally
quantified by $D_{n}(4000)$ \citep{ba99}, which measures the ratio of
the average flux density $F_{\nu}$ in continuum regions blueward
(3850-3950 \AA) and redward (4000-4100 \AA) of the break, or
by \dn\ \citep{br83}, which uses larger continuum regions around the
break (3750-3950 \AA\ and 4050-4250 \AA). We measure \dn\ for all our
composite SEDs, as it better matches the resolution of our composite
SEDs. We determine the confidence intervals using Monte Carlo
simulations. For the photometric errors we take the error on the mean,
as explained in Section~\ref{construct}.

\begin{deluxetable*}{l l l l c c c c c c c c c}
  \tabletypesize{\scriptsize}  \tablecaption{Stellar population
    properties \label{tab:mod}} \tablewidth{0pt} \tablehead{  & & &
    & \colhead{Log\,$t$} & \colhead{Log\,$\tau$} & \colhead{\av} &
    &  \colhead{Log SSFR} & & & & \colhead{$W_{\rm
    H\alpha+}$} \\  \colhead{ID}  & \colhead{$N$}
    & \colhead{IT\tablenotemark{a}} & \colhead{Model} &  \colhead{yr}
    & \colhead{yr} & \colhead{mag} & \colhead{$Z$}
    &  \colhead{yr$^{-1}$} & \colhead{$\chi^2_{\rm red}$}
    &  \colhead{\dn} & \colhead{\dn$_{\rm corr}$}
    &  \colhead{\AA}}     \startdata      1 & 123 &  14 &  BC03 &    9.60$^{+    0.11}_{-    0.10}$ &    8.60$^{+    0.13}_{-    0.15}$ &    0.40$^{+    0.10}_{-    0.14}$ &     0.020 &    -11.79$^{+      0.11}_{-      0.38}$ &    0.22 &    1.76$^{+    0.02}_{-    0.02}$ &    1.95$^{+    0.11}_{-    0.09}$ &    14.4 $^{+    10.0 }_{-    10.2}$ \\
     &     &     &   M05 &    9.50$^{+    0.11}_{-    0.06}$ &    8.40$^{+    0.23}_{-    0.89}$ &    0.60$^{+    0.04}_{-    0.09}$ &     0.010 &    -12.60$^{+      1.22}_{-     23.17}$ &    0.38 & & &\\
   2 & 314 &   2 &  BC03 &    9.60$^{+    0.11}_{-    0.39}$ &    8.60$^{+    0.15}_{-    1.60}$ &    0.20$^{+    0.53}_{-    0.15}$ &     0.020 &    -11.79$^{+      0.19}_{-    \infty}$ &    0.27 &    1.75$^{+    0.01}_{-    0.01}$ &    1.94$^{+    0.10}_{-    0.09}$ &    14.4 $^{+     9.9 }_{-     8.8}$ \\
     &     &     &   M05 &    9.50$^{+    0.10}_{-    0.06}$ &    8.40$^{+    0.20}_{-    1.05}$ &    0.40$^{+    0.03}_{-    0.06}$ &     0.010 &    -12.60$^{+      0.80}_{-     86.40}$ &    0.36 & & &\\
   3 & 455 &   1 &  BC03 &    9.25$^{+    0.36}_{-    0.02}$ &    7.60$^{+    1.01}_{-    0.39}$ &    0.50$^{+    0.03}_{-    0.50}$ &     0.008 &    -24.93$^{+     13.25}_{-    \infty}$ &    0.19 &    1.69$^{+    0.01}_{-    0.01}$ &    1.81$^{+    0.10}_{-    0.09}$ &     6.2 $^{+     9.8 }_{-     9.3}$ \\
     &     &     &   M05 &    9.60$^{+    0.04}_{-    0.10}$ &    8.60$^{+    0.08}_{-    0.20}$ &    0.20$^{+    0.02}_{-    0.09}$ &     0.010 &    -11.80$^{+      0.00}_{-      0.80}$ &    0.28 & & &\\
   4 & 220 &   5 &  BC03 &    9.25$^{+    0.12}_{-    0.02}$ &    7.80$^{+    0.56}_{-    0.37}$ &    0.30$^{+    0.04}_{-    0.20}$ &     0.008 &    -18.35$^{+      6.84}_{-     11.24}$ &    0.17 &    1.64$^{+    0.01}_{-    0.01}$ &    1.77$^{+    0.10}_{-    0.08}$ &     2.1 $^{+     8.5 }_{-     7.5}$ \\
     &     &     &   M05 &    9.60$^{+    0.01}_{-    0.11}$ &    8.60$^{+    0.01}_{-    0.13}$ &    0.00$^{+    0.01}_{-    0.00}$ &     0.010 &    -11.80$^{+      0.11}_{-      0.38}$ &    0.36 & & &\\
   5 & 100 &  22 &  BC03 &    9.35$^{+    0.15}_{-    0.02}$ &    8.50$^{+    0.10}_{-    0.06}$ &    0.70$^{+    0.06}_{-    0.30}$ &     0.020 &    -10.59$^{+      0.00}_{-      0.41}$ &    0.26 &    1.61$^{+    0.01}_{-    0.02}$ &    1.70$^{+    0.10}_{-    0.08}$ &    24.6 $^{+     9.9 }_{-     9.7}$ \\
     &     &     &   M05 &    9.55$^{+    0.01}_{-    0.05}$ &    8.70$^{+    0.04}_{-    0.10}$ &    0.40$^{+    0.02}_{-    0.10}$ &     0.010 &    -10.79$^{+      0.00}_{-      0.22}$ &    0.53 & & &\\
   6 &  22 &  32 &  BC03 &    9.05$^{+    0.03}_{-    0.21}$ &    8.00$^{+    0.06}_{-    1.00}$ &    0.10$^{+    0.33}_{-    0.04}$ &     0.020 &    -11.67$^{+      0.47}_{-     22.11}$ &    0.34 &    1.54$^{+    0.03}_{-    0.04}$ &    1.67$^{+    0.12}_{-    0.06}$ &    19.1 $^{+    15.7 }_{-    17.3}$ \\
     &     &     &   M05 &    8.80$^{+    0.00}_{-    0.00}$ &    7.00$^{+    0.00}_{-    0.00}$ &    0.00$^{+    0.00}_{-    0.00}$ &     0.020 &    -30.81$^{+      0.00}_{-      0.00}$ &    3.02 & & &\\
   7 &  75 &  16 &  BC03 &    9.10$^{+    0.10}_{-    0.15}$ &    8.10$^{+    0.10}_{-    1.10}$ &    0.20$^{+    0.41}_{-    0.10}$ &     0.020 &    -11.29$^{+      0.09}_{-     29.59}$ &    0.22 &    1.55$^{+    0.02}_{-    0.02}$ &    1.66$^{+    0.10}_{-    0.07}$ &    20.5 $^{+     9.6 }_{-     9.3}$ \\
     &     &     &   M05 &    8.85$^{+    0.00}_{-    0.00}$ &    7.00$^{+    0.00}_{-    0.00}$ &    0.00$^{+    0.00}_{-    0.00}$ &     0.020 &    -33.79$^{+      0.00}_{-      0.00}$ &    1.20 & & &\\
   8 &  69 &  30 &  BC03 &    9.25$^{+    0.11}_{-    0.12}$ &    8.40$^{+    0.12}_{-    0.13}$ &    1.20$^{+    0.11}_{-    0.32}$ &     0.008 &    -10.49$^{+      0.10}_{-      0.10}$ &    0.47 &    1.58$^{+    0.02}_{-    0.02}$ &    1.66$^{+    0.10}_{-    0.07}$ &    38.7 $^{+     8.9 }_{-    10.1}$ \\
     &     &     &   M05 &    8.20$^{+    0.11}_{-    0.10}$ &    7.10$^{+    0.12}_{-    0.10}$ &    2.30$^{+    0.10}_{-    0.11}$ &     0.020 &    -11.34$^{+      0.09}_{-      1.93}$ &    0.58 & & &\\
   9 &  99 &   8 &  BC03 &    9.35$^{+    0.03}_{-    0.11}$ &    8.50$^{+    0.06}_{-    0.11}$ &    1.10$^{+    0.30}_{-    0.04}$ &     0.020 &    -10.59$^{+      0.29}_{-      0.00}$ &    0.48 &    1.56$^{+    0.02}_{-    0.01}$ &    1.65$^{+    0.09}_{-    0.07}$ &    42.7 $^{+    10.6 }_{-    10.1}$ \\
     &     &     &   M05 &    8.15$^{+    0.26}_{-    0.11}$ &    7.00$^{+    0.51}_{-    0.00}$ &    2.50$^{+    0.11}_{-    0.40}$ &     0.020 &    -11.84$^{+      1.89}_{-      1.43}$ &    0.63 & & &\\
  10 &  65 &  25 &  BC03 &    9.35$^{+    0.01}_{-    0.10}$ &    8.50$^{+    0.09}_{-    0.10}$ &    0.50$^{+    0.10}_{-    0.09}$ &     0.008 &    -10.58$^{+      0.09}_{-      0.00}$ &    0.23 &    1.54$^{+    0.03}_{-    0.03}$ &    1.62$^{+    0.10}_{-    0.06}$ &    33.1 $^{+    10.3 }_{-     8.5}$ \\
     &     &     &   M05 &    9.30$^{+    0.01}_{-    0.10}$ &    8.40$^{+    0.03}_{-    0.10}$ &    0.00$^{+    0.10}_{-    0.00}$ &     0.020 &    -10.83$^{+      0.11}_{-      0.00}$ &    0.37 & & &\\
  11 & 101 &  12 &  BC03 &    9.40$^{+    0.11}_{-    0.11}$ &    8.60$^{+    0.12}_{-    0.13}$ &    0.70$^{+    0.13}_{-    0.12}$ &     0.008 &    -10.40$^{+      0.10}_{-      0.09}$ &    0.34 &    1.51$^{+    0.02}_{-    0.01}$ &    1.58$^{+    0.09}_{-    0.07}$ &    39.1 $^{+    10.1 }_{-     8.8}$ \\
     &     &     &   M05 &    9.45$^{+    0.04}_{-    1.46}$ &    8.60$^{+    0.09}_{-    1.60}$ &    0.10$^{+    1.93}_{-    0.03}$ &     0.010 &    -10.69$^{+      1.03}_{-      2.58}$ &    0.59 & & &\\
  12 &  53 &  27 &  BC03 &    9.35$^{+    0.15}_{-    0.11}$ &    8.50$^{+    0.13}_{-    0.13}$ &    1.40$^{+    0.10}_{-    0.50}$ &     0.008 &    -10.58$^{+      0.09}_{-      0.42}$ &    1.05 &    1.50$^{+    0.02}_{-    0.02}$ &    1.58$^{+    0.09}_{-    0.07}$ &    17.7 $^{+    10.4 }_{-    11.0}$ \\
     &     &     &   M05 &    8.35$^{+    0.65}_{-    0.17}$ &    7.10$^{+    0.90}_{-    0.10}$ &    2.40$^{+    0.23}_{-    1.90}$ &     0.020 &    -13.37$^{+      2.74}_{-      2.78}$ &    1.08 & & &\\
  13 & 117 &  10 &  BC03 &    8.05$^{+    0.05}_{-    0.01}$ &    7.00$^{+    0.10}_{-    0.00}$ &    2.40$^{+    0.02}_{-    0.04}$ &     0.008 &    -10.71$^{+      0.37}_{-      0.00}$ &    0.61 &    1.51$^{+    0.01}_{-    0.01}$ &    1.45$^{+    0.08}_{-    0.07}$ &    63.3 $^{+    10.3 }_{-    10.2}$ \\
     &     &     &   M05 &    8.10$^{+    0.15}_{-    0.09}$ &    7.30$^{+    0.10}_{-    0.20}$ &    2.30$^{+    0.02}_{-    0.33}$ &     0.020 &     -9.16$^{+      0.00}_{-      1.18}$ &    0.51 & & &\\
  14 &  77 &  19 &  BC03 &    8.10$^{+    0.03}_{-    0.10}$ &    7.20$^{+    0.07}_{-    0.20}$ &    2.60$^{+    0.10}_{-    0.04}$ &     0.008 &     -9.66$^{+      0.00}_{-      1.05}$ &    0.49 &    1.49$^{+    0.01}_{-    0.01}$ &    1.45$^{+    0.08}_{-    0.07}$ &    69.4 $^{+    11.4 }_{-    11.3}$ \\
     &     &     &   M05 &    8.15$^{+    0.05}_{-    0.30}$ &    7.40$^{+    0.12}_{-    0.40}$ &    2.50$^{+    0.11}_{-    0.21}$ &     0.020 &     -9.01$^{+      0.12}_{-      0.92}$ &    0.44 & & &\\
  15 &  66 &  15 &  BC03 &    8.10$^{+    0.00}_{-    0.06}$ &    7.10$^{+    0.01}_{-    0.10}$ &    2.80$^{+    0.10}_{-    0.00}$ &     0.008 &    -10.34$^{+      0.42}_{-      0.37}$ &    0.88 &    1.43$^{+    0.02}_{-    0.02}$ &    1.40$^{+    0.08}_{-    0.06}$ &    61.7 $^{+    11.6 }_{-    11.0}$ \\
     &     &     &   M05 &    7.90$^{+    0.38}_{-    0.04}$ &    7.00$^{+    0.56}_{-    0.00}$ &    2.80$^{+    0.03}_{-    0.40}$ &     0.020 &     -9.47$^{+      0.41}_{-      1.24}$ &    0.81 & & &\\
  16 & 157 &   6 &  BC03 &    8.05$^{+    0.01}_{-    0.05}$ &    7.10$^{+    0.02}_{-    0.10}$ &    2.10$^{+    0.00}_{-    0.01}$ &     0.008 &     -9.92$^{+      0.00}_{-      0.32}$ &    0.59 &    1.44$^{+    0.01}_{-    0.01}$ &    1.40$^{+    0.08}_{-    0.07}$ &    67.4 $^{+    10.2 }_{-     9.8}$ \\
     &     &     &   M05 &    8.00$^{+    0.30}_{-    0.05}$ &    7.20$^{+    0.42}_{-    0.11}$ &    2.00$^{+    0.02}_{-    0.12}$ &     0.020 &     -9.06$^{+      0.15}_{-      0.19}$ &    0.49 & & &\\
  17 & 103 &  17 &  BC03 &    8.10$^{+    1.15}_{-    0.07}$ &    7.20$^{+    1.40}_{-    0.20}$ &    1.90$^{+    0.03}_{-    1.00}$ &     0.008 &     -9.66$^{+      0.00}_{-      1.05}$ &    0.72 &    1.44$^{+    0.01}_{-    0.01}$ &    1.40$^{+    0.08}_{-    0.06}$ &    66.2 $^{+     8.6 }_{-    10.2}$ \\
     &     &     &   M05 &    8.15$^{+    0.15}_{-    0.23}$ &    7.40$^{+    0.11}_{-    0.35}$ &    1.80$^{+    0.05}_{-    0.31}$ &     0.020 &     -9.01$^{+      0.00}_{-      0.53}$ &    0.56 & & &\\
  18 &  96 &  13 &  BC03 &    8.05$^{+    0.00}_{-    0.00}$ &    7.20$^{+    0.00}_{-    0.00}$ &    2.30$^{+    0.00}_{-    0.00}$ &     0.008 &     -9.34$^{+      0.00}_{-      0.00}$ &    0.50 &    1.43$^{+    0.01}_{-    0.01}$ &    1.38$^{+    0.08}_{-    0.06}$ &    86.0 $^{+    11.5 }_{-    10.2}$ \\
     &     &     &   M05 &    7.95$^{+    0.20}_{-    0.06}$ &    7.20$^{+    0.30}_{-    0.12}$ &    2.20$^{+    0.02}_{-    0.05}$ &     0.020 &     -8.82$^{+      0.13}_{-      0.15}$ &    0.49 & & &\\
  19 &  36 &  29 &  BC03 &    8.10$^{+    0.08}_{-    0.20}$ &    7.20$^{+    0.24}_{-    0.20}$ &    2.60$^{+    0.30}_{-    0.09}$ &     0.020 &     -9.66$^{+      0.88}_{-      1.05}$ &    0.46 &    1.40$^{+    0.02}_{-    0.02}$ &    1.37$^{+    0.08}_{-    0.06}$ &    86.1 $^{+    17.7 }_{-    20.7}$ \\
     &     &     &   M05 &    7.90$^{+    0.02}_{-    0.15}$ &    7.20$^{+    0.05}_{-    0.20}$ &    2.80$^{+    0.00}_{-    0.02}$ &     0.020 &     -8.60$^{+      0.00}_{-      0.12}$ &    0.56 & & &\\
  20 &  35 &  26 &  BC03 &    7.95$^{+    0.40}_{-    0.06}$ &    7.20$^{+    0.50}_{-    0.13}$ &    2.60$^{+    0.03}_{-    0.36}$ &     0.008 &     -8.81$^{+      0.12}_{-      0.29}$ &    0.55 &    1.43$^{+    0.02}_{-    0.02}$ &    1.37$^{+    0.08}_{-    0.06}$ &   108.0 $^{+    13.5 }_{-    14.1}$ \\
     &     &     &   M05 &    8.00$^{+    0.30}_{-    0.22}$ &    7.40$^{+    0.51}_{-    0.40}$ &    2.50$^{+    0.02}_{-    0.20}$ &     0.020 &     -8.43$^{+      0.10}_{-      0.72}$ &    0.65 & & &\\
  21 & 140 &   4 &  BC03 &    8.10$^{+    0.41}_{-    0.16}$ &    7.30$^{+    0.62}_{-    0.30}$ &    1.90$^{+    0.06}_{-    0.23}$ &     0.008 &     -9.15$^{+      0.27}_{-      0.77}$ &    0.62 &    1.39$^{+    0.01}_{-    0.01}$ &    1.35$^{+    0.07}_{-    0.06}$ &    79.5 $^{+     9.6 }_{-    10.1}$ \\
     &     &     &   M05 &    8.00$^{+    0.16}_{-    0.05}$ &    7.30$^{+    0.23}_{-    0.10}$ &    1.80$^{+    0.01}_{-    0.05}$ &     0.020 &     -8.69$^{+      0.00}_{-      0.13}$ &    0.51 & & &\\
  22 &  82 &  23 &  BC03 &    8.00$^{+    0.05}_{-    0.00}$ &    7.00$^{+    0.10}_{-    0.00}$ &    1.70$^{+    0.00}_{-    0.01}$ &     0.008 &    -10.24$^{+      0.32}_{-      0.00}$ &    0.59 &    1.38$^{+    0.01}_{-    0.01}$ &    1.34$^{+    0.07}_{-    0.06}$ &    66.9 $^{+     9.7 }_{-     9.4}$ \\
     &     &     &   M05 &    8.00$^{+    0.05}_{-    0.02}$ &    7.20$^{+    0.10}_{-    0.04}$ &    1.60$^{+    0.02}_{-    0.04}$ &     0.020 &     -9.06$^{+      0.15}_{-      0.00}$ &    0.51 & & &\\
  23 &  52 &  20 &  BC03 &    8.00$^{+    0.46}_{-    0.05}$ &    7.30$^{+    0.72}_{-    0.23}$ &    2.20$^{+    0.02}_{-    0.31}$ &     0.008 &     -8.69$^{+      0.11}_{-      0.87}$ &    0.32 &    1.37$^{+    0.02}_{-    0.01}$ &    1.33$^{+    0.08}_{-    0.06}$ &    92.4 $^{+    11.4 }_{-    11.1}$ \\
     &     &     &   M05 &    7.85$^{+    0.56}_{-    0.05}$ &    7.00$^{+    1.01}_{-    0.00}$ &    1.90$^{+    0.20}_{-    0.01}$ &     0.010 &     -9.15$^{+      0.78}_{-      0.00}$ &    0.40 & & &\\
  24 &  26 &  31 &  BC03 &    8.20$^{+    0.06}_{-    0.20}$ &    7.40$^{+    0.12}_{-    0.40}$ &    2.10$^{+    0.03}_{-    0.02}$ &     0.020 &     -9.25$^{+      0.15}_{-      0.99}$ &    0.44 &    1.35$^{+    0.02}_{-    0.02}$ &    1.30$^{+    0.08}_{-    0.06}$ &   130.7 $^{+    14.1 }_{-    17.7}$ \\
     &     &     &   M05 &    8.10$^{+    0.03}_{-    0.36}$ &    7.30$^{+    0.07}_{-    0.30}$ &    1.90$^{+    0.42}_{-    0.10}$ &     0.040 &     -9.16$^{+      0.66}_{-      0.67}$ &    0.63 & & &\\
  25 & 159 &   3 &  BC03 &    8.05$^{+    0.45}_{-    0.10}$ &    7.00$^{+    0.90}_{-    0.00}$ &    1.40$^{+    0.31}_{-    0.01}$ &     0.020 &    -10.71$^{+      1.81}_{-      0.00}$ &    0.49 &    1.34$^{+    0.01}_{-    0.01}$ &    1.30$^{+    0.07}_{-    0.06}$ &    94.1 $^{+    10.6 }_{-     9.5}$ \\
     &     &     &   M05 &    8.00$^{+    0.30}_{-    0.02}$ &    7.30$^{+    0.40}_{-    0.03}$ &    1.60$^{+    0.02}_{-    0.11}$ &     0.020 &     -8.69$^{+      0.09}_{-      0.02}$ &    0.45 & & &\\
  26 &  80 &  24 &  BC03 &    7.95$^{+    0.50}_{-    0.05}$ &    7.20$^{+    0.80}_{-    0.20}$ &    1.90$^{+    0.03}_{-    0.30}$ &     0.008 &     -8.81$^{+      0.23}_{-      0.75}$ &    0.41 &    1.33$^{+    0.01}_{-    0.01}$ &    1.30$^{+    0.07}_{-    0.06}$ &   107.1 $^{+     9.3 }_{-    11.3}$ \\
     &     &     &   M05 &    8.00$^{+    0.05}_{-    0.22}$ &    7.40$^{+    0.10}_{-    0.40}$ &    1.80$^{+    0.01}_{-    0.20}$ &     0.020 &     -8.43$^{+      0.06}_{-      0.72}$ &    0.50 & & &\\
 27 & 108 &  18 &  BC03 &    8.10$^{+    0.48}_{-    0.10}$ &    7.40$^{+    0.76}_{-    0.21}$ &    1.60$^{+    0.10}_{-    0.12}$ &     0.008 &     -8.78$^{+      0.10}_{-      0.28}$ &    0.55 &    1.31$^{+    0.01}_{-    0.01}$ &    1.27$^{+    0.07}_{-    0.06}$ &   105.8 $^{+    10.3 }_{-    10.5}$ \\
     &     &     &   M05 &    8.10$^{+    0.20}_{-    0.25}$ &    7.50$^{+    0.31}_{-    0.50}$ &    1.50$^{+    0.05}_{-    0.10}$ &     0.020 &     -8.52$^{+      0.00}_{-      0.63}$ &    0.61 & & &\\
 \enddata
\end{deluxetable*}

\begin{deluxetable*}{l l l l c c c c c c c c c}
  \tabletypesize{\scriptsize}  \tablecaption{Stellar population
    properties,
    continued \label{tab:mod}} \tablewidth{0pt} \tablehead{  & & &
    & \colhead{Log\,$t$} & \colhead{Log\,$\tau$} & \colhead{\av} &
    &  \colhead{Log SSFR} & & & & \colhead{$W_{\rm
    H\alpha+}$}\\  \colhead{ID}  & \colhead{$N$}
    & \colhead{IT\tablenotemark{a}} & \colhead{Model} &  \colhead{yr}
    & \colhead{yr} & \colhead{mag} & \colhead{$Z$}
    &  \colhead{yr$^{-1}$} & \colhead{$\chi^2_{\rm red}$}
    &  \colhead{\dn} & \colhead{\dn$_{\rm corr}$}
    &  \colhead{\AA}}     \startdata     28 & 144 &   7 &  BC03 &    8.05$^{+    0.31}_{-    0.05}$ &    7.10$^{+    0.63}_{-    0.10}$ &    1.20$^{+    0.30}_{-    0.03}$ &     0.020 &     -9.92$^{+      1.14}_{-      0.32}$ &    0.49 &    1.30$^{+    0.01}_{-    0.01}$ &    1.25$^{+    0.07}_{-    0.06}$ &   106.2 $^{+    11.4 }_{-    10.5}$ \\
     &     &     &   M05 &    8.00$^{+    0.20}_{-    0.10}$ &    7.30$^{+    0.30}_{-    0.10}$ &    1.30$^{+    0.10}_{-    0.04}$ &     0.020 &     -8.69$^{+      0.09}_{-      0.00}$ &    0.49 & & &\\
  29 &  64 &  28 &  BC03 &    8.20$^{+    0.06}_{-    0.15}$ &    7.60$^{+    0.12}_{-    0.20}$ &    1.40$^{+    0.11}_{-    0.03}$ &     0.008 &     -8.61$^{+      0.09}_{-      0.00}$ &    0.54 &    1.26$^{+    0.01}_{-    0.01}$ &    1.22$^{+    0.07}_{-    0.06}$ &   102.8 $^{+    11.5 }_{-    12.5}$ \\
     &     &     &   M05 &    8.15$^{+    0.05}_{-    0.00}$ &    7.60$^{+    0.10}_{-    0.01}$ &    1.30$^{+    0.01}_{-    0.01}$ &     0.020 &     -8.47$^{+      0.05}_{-      0.00}$ &    0.61 & & &\\
  30 & 142 &   9 &  BC03 &    8.35$^{+    0.15}_{-    0.43}$ &    7.80$^{+    0.20}_{-    0.76}$ &    1.20$^{+    0.06}_{-    0.12}$ &     0.008 &     -8.65$^{+      0.04}_{-      0.59}$ &    0.45 &    1.24$^{+    0.01}_{-    0.01}$ &    1.21$^{+    0.06}_{-    0.06}$ &   104.0 $^{+     9.0 }_{-     9.7}$ \\
     &     &     &   M05 &    8.05$^{+    0.25}_{-    0.10}$ &    7.40$^{+    0.39}_{-    0.10}$ &    1.10$^{+    0.10}_{-    0.04}$ &     0.020 &     -8.60$^{+      0.10}_{-      0.00}$ &    0.48 & & &\\
  31 & 101 &  11 &  BC03 &    8.35$^{+    0.07}_{-    0.18}$ &    7.80$^{+    0.14}_{-    0.26}$ &    1.00$^{+    0.12}_{-    0.06}$ &     0.008 &     -8.65$^{+      0.10}_{-      0.06}$ &    0.50 &    1.22$^{+    0.01}_{-    0.01}$ &    1.18$^{+    0.06}_{-    0.06}$ &   109.7 $^{+    12.9 }_{-    11.6}$ \\
     &     &     &   M05 &    8.00$^{+    0.01}_{-    0.05}$ &    7.40$^{+    0.01}_{-    0.10}$ &    1.00$^{+    0.01}_{-    0.03}$ &     0.020 &     -8.43$^{+      0.00}_{-      0.07}$ &    0.55 & & &\\
  32 &  50 &  21 &  BC03 &    8.30$^{+    0.26}_{-    0.18}$ &    7.80$^{+    0.52}_{-    0.25}$ &    0.90$^{+    0.11}_{-    0.05}$ &     0.008 &     -8.51$^{+      0.12}_{-      0.04}$ &    0.55 &    1.14$^{+    0.01}_{-    0.01}$ &    1.13$^{+    0.06}_{-    0.05}$ &   138.5 $^{+    13.5 }_{-    10.1}$ \\
     &     &     &   M05 &    8.00$^{+    0.28}_{-    0.05}$ &    7.40$^{+    0.47}_{-    0.03}$ &    0.80$^{+    0.10}_{-    0.02}$ &     0.020 &     -8.43$^{+      0.16}_{-      0.00}$ &    0.63 & & &\\
 \enddata

  \tablecomments{We fit a grid of stellar population models assuming a
    delayed exponential SFH ($\psi(t)\propto t \,{\rm exp}(-t/\tau)$)
    with log\,($t$/yr) between 7.6 and 10.1, in steps of 0.05,
    log\,($\tau$/yr) between 7 and 10 in steps of 0.1, and \av\
    between 0 and 3 in steps of 0.1\,mag.  We assume the \cite{ca00}
    attenuation curve and a \cite{sa55} IMF. For each model we explore
    three different metallicities: subsolar (0.01 and 0.008 for
    the \cite{ma05} and \cite{bc03} models, respectively), solar, and
    supersolar (0.04 and 0.05 for the \cite{ma05} and \cite{bc03}
    models, respectively). The errors represent the 68\% confidence
    intervals.}  

  \tablenotetext{a}{The iteration in which the SED is constructed}

\end{deluxetable*}

\begin{figure*} 
  \includegraphics[width=0.98\textwidth]{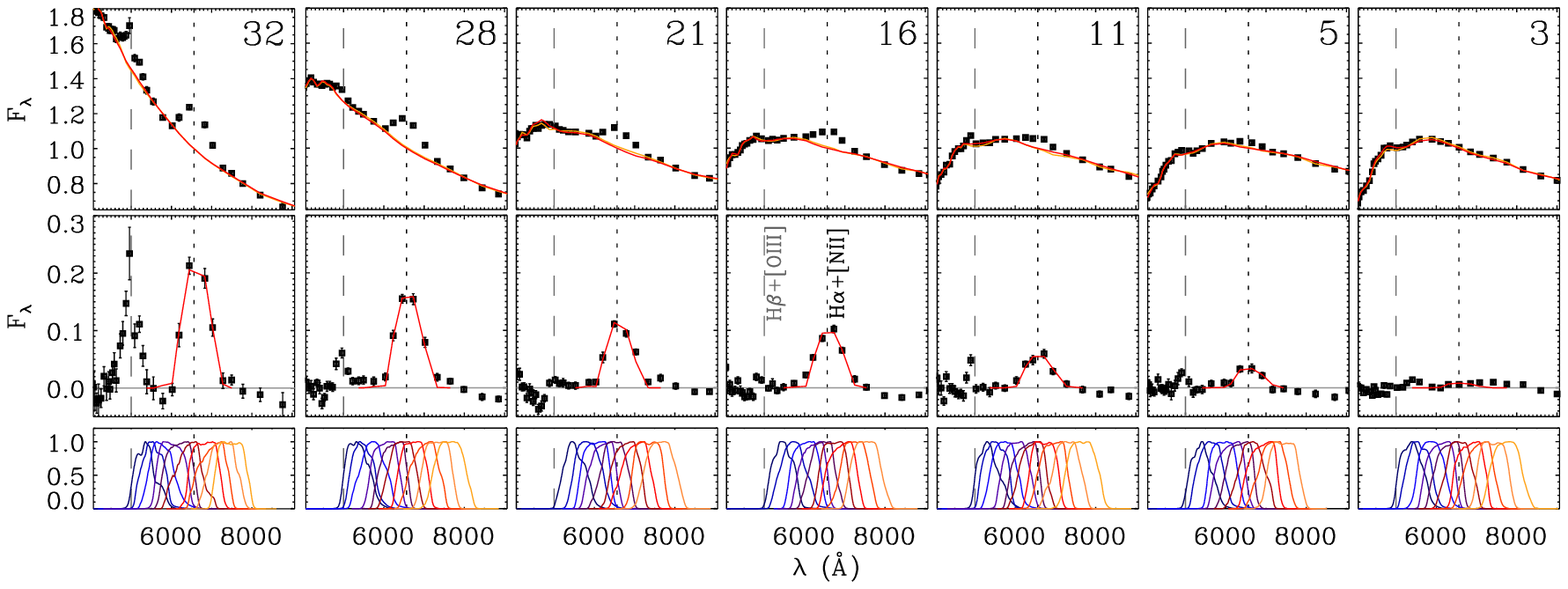}

  \caption{Illustration of H$\alpha$ measurements. The presented
     composite SEDs cover a wide range in \wha, and are ordered from
     the highest to the lowest \wha\ when going from left to right.
     The continuum emission is fit locally, between 4000-10000~\AA,
     while masking $H\beta$, [O\,{\sc iii}], $H\alpha$, \nt,
     and \st. The red and orange curves represent the \cite{bc03}
     and \cite{ma05} continuum models, respectively. In the middle
     panels we remove the \cite{bc03} continuum model and fit the
     remaining emission line (red), taking into account the composite
     filter curves, as represented in the bottom panels. No redshift
     uncertainties are assumed when fitting the line. The consistency
     between the observed shapes and those predicted by the filter
     curves demonstrate the accuracy of our photometric redshifts (see
     Figure~\ref{hadz}). The middle panels also show indications for
     H$\beta$ and/or [O\,{\sc iii}] at $\sim$5000\,\AA\ in certain
     cases (dashed lines).\label{ha}}
  
\end{figure*}

The break measurements are affected by the low resolution of the
medium bands and possible photometric redshifts errors. We correct for
the resolution as explained in detail in Appendix C. Summarized, we
derive a correction factor for each break measurement using the
difference between the break measurement for the original best-fit
stellar population model and the best-fit model convolved by the
composite filter curves (see Figure~\label{dncorr} in Appendix c). The
exact correction factor depends primarily on the strength of the
break, but also on the sampling of the composite filters. The original
and corrected values for \dn\ are given in Table 1.

Correcting for photometric redshift errors is more challenging, as we
do not know the uncertainties for the different SED types. In appendix
B we show the effects of the break measurement in case the median
random scatter is $\Delta z/(1+z) =$ 0.005, 0.01, 0.02, or 0.04. We do
not apply this correction, but caution that the break measurements, in
particular for large values of \dn, are possibly underestimated.

\begin{figure} 
  \begin{center} 
    \includegraphics[width=0.48\textwidth]{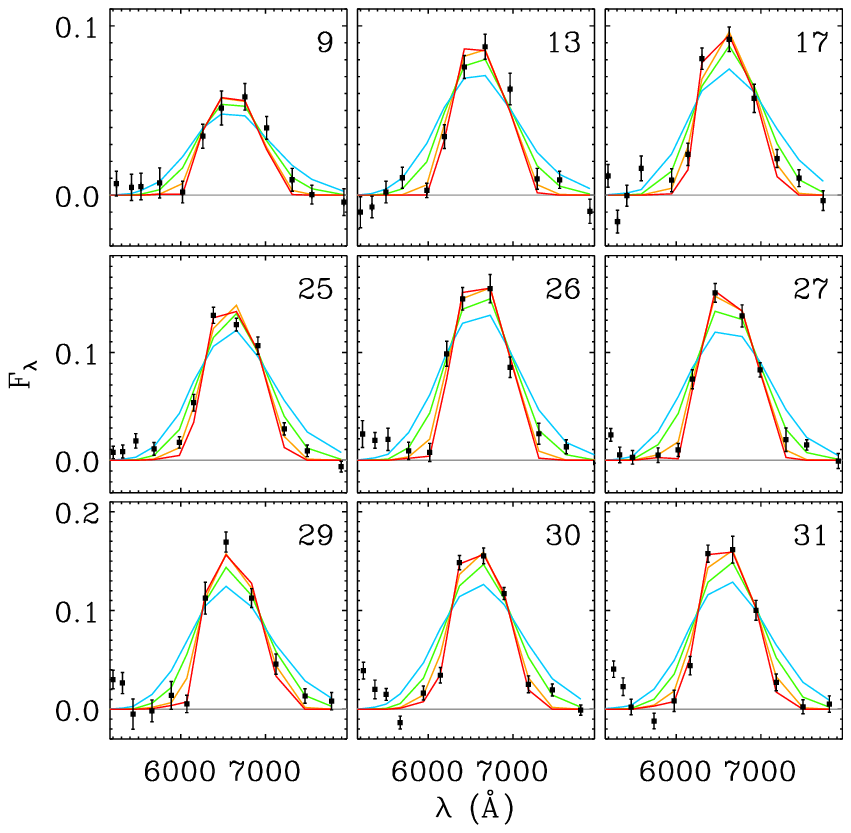}
    \caption{Fits to \ha\ emission lines for different photometric
      redshift uncertainties. The red curve is the best-fit model when
      assuming that the photometric redshifts are perfect. For the
      orange, green, and blue models the random scatter $\Delta
      z/(1+z)$ is 0.02, 0.04, and 0.06, respectively. This plot
      illustrates that the NMBS photometric redshifts are certain to
      $\Delta z/(1+z) \lesssim 0.02$. \label{hadz}}
  \end{center}
\end{figure}

\subsection{\ha\ emission line}

Remarkably, we see a bump at the location of \ha\ in  most SEDs. In
several composite SEDs we also see indications for H$\beta$ and/or
[O\,{\sc iii}] near 0.5\,$\mu$m. We measure the equivalent width (EW)
of \ha+\nt+\st\ in the following way. As the best stellar population
fits to full SED do not always provide a good fit to the continuum
region around \ha, we refit the SED restricting the wavelength region
to $4000<\lambda<10000$\, \AA, while masking the \ha, \nt, \st,
H$\beta$, and [O\,{\sc iii}] emission lines. We subtract the continuum
model, thus correcting for the underlying \ha\ absorption. Next, we
make an emission-line model, assuming a fixed ratio of the [N\,{\sc
ii} $\lambda$6583]/\ha\ of 0.25, [N\,{\sc ii} $\lambda$6583]/[N\,{\sc
ii} $\lambda$6548] of 3, and [S\,{\sc ii}]/\ha\ of 0.3. We convolve
the model with the composite filter curves, scale it to fit the data,
and derive the total EW of the blended \ha, \nt, and \st\ lines. 

Confidence intervals on \wha\ are determined using Monte Carlo
simulations. We simultaneously vary the photometric uncertainties
(following the error on the mean, see Section~\ref{construct}) and the
continuum level. For the latter we assume an uncertainty of 1\%. We
also randomly vary the emission-line ratios in the simulations,
assuming an uncertainty of 0.1 on both [N\,{\sc ii}
$\lambda$6583]/\ha\ and [S\,{\sc ii}]/\ha. However, as we derive the
total EWs, the line ratios will barely affect the best-fit value or
its confidence intervals.

Figure~\ref{ha} shows examples of the \ha\ emission line for seven
composite SEDs, ordered by decreasing \wha. The consistency between
the shape of the observed emission lines and the convolved
emission-line models is striking. This result demonstrates the
accuracy of our photometric redshifts. In Figure~\ref{hadz} we assess
the influence of different values for $\Delta z/(1+z)$ on the shape
and strength of the detected \ha\ emission line. The observed shapes
give us a rough constraint on the photometric redshift uncertainties
for certain SED types. The average random uncertainty is $\Delta
z/(1+z) \lesssim 0.02$, consistent with the uncertainties found
by \cite{wh11} when comparing photometric with spectroscopic
redshifts. We cannot rule out catastrophic outliers, as they simply
would increase the apparent variation among individual SEDs. Thus, the
EWs of \ha\ might be slightly underestimated.

\subsection{Rest-frame optical morphologies}

For each SED type we also show the morphologies of a subsample of the
galaxies. We choose the nine galaxies that are closest in redshift to
0.9, as for these galaxies the available HST/ACS i-band imaging from
COSMOS probes the rest-frame optical. The images are displayed in
Figure~\ref{compilation} next to the composite SEDs. In this paper we
will not give a quantitative analysis of the sizes and morphologies;
this will be discussed in detail in a future paper.

\begin{figure*}  
  \begin{center}   
    \includegraphics[width=0.98\textwidth]{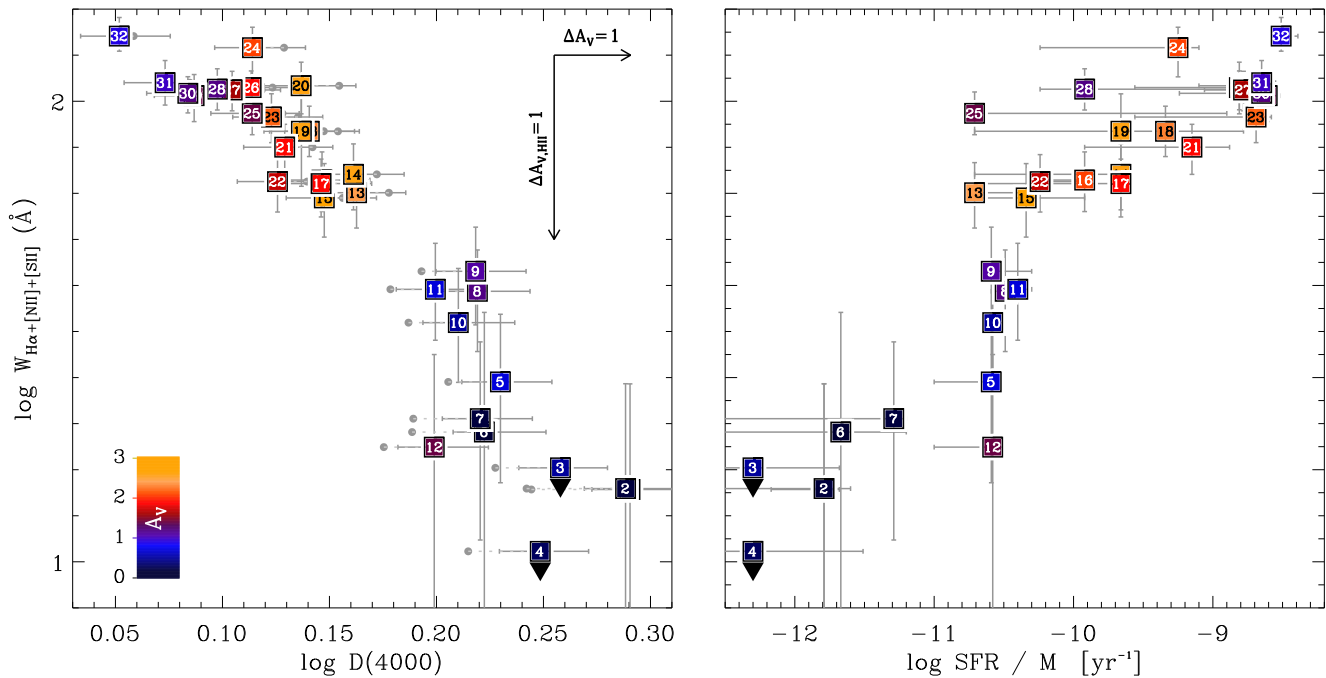}

    \caption{Left: Correlation between \wha\ and \dn\ for the 32
      composite SEDs. The corresponding IDs are printed in each
      symbol. The color coding of the symbols reflect their best-fit
      value for \av, as indicated by the color bar in the bottom
      left. The triangles represent 1$\sigma$ upper limits. The gray
      dots represent the values for \dn\ which are uncorrected for the
      NMBS resolution (see Appendix B). They are connected to the
      corrected values by the dotted gray lines. The horizontal dust
      vectors indicates the total attenuation. The vertical dust
      vector indicates extra extinction towards H\,{\sc ii}
      regions. Right: Correlation between \wha\ and the best-fit
      SSFR. This figure illustrates that \wha, \dn, and the best-fit
      SSFR are strongly correlated: galaxies with shallow breaks have
      high values for \wha\ and are best fit by stellar population
      models with high SSFR, while older galaxies with low SSFRs and
      strong 4000~\AA\ breaks have low values for \wha. \label{corr}}  

  \end{center} 
\end{figure*}

\section{CORRELATIONS}

\subsection{Star Formation History Indicators}

In the previous section we have directly measured two key indicators
of present and past star formation in galaxies: \dn, which is
sensitive to the age of the stellar population (as well as
metallicity), and \ha, which measures the instantaneous SFR normalized
to the amount of stellar continuum emission at 6563\,\AA.  Together,
these parameters measure the star formation history of a galaxy at
different times, with \dn\ mostly sensitive to past star formation
and \ha\ mostly sensitive to current star formation. Additionally, we
derive the SSFR by comparing the full continuum shape with SPS
models. The SSFR is the SFR divided by the stellar mass, and thus,
similar to \wha, it is a measure of the current versus past star
formation, though independently determined. 

In Figure~\ref{corr} we compare the different parameters and
show \wha\ versus both \dn\ and the SSFR. Note that \dn\ and the
modeled SSFRs are not completely independent, as both are derived from
the continuum shape, and thus are not directly compared. Remarkably,
both panels show a tight correlation. Galaxies with weak breaks have
high values for \wha\ and are best fit by stellar population models
with high SSFR, while galaxies with low SSFRs and strong 4000~\AA\
breaks have low values for \wha. This is a remarkable result given
that the relations are derived without spectroscopy, and are solely
based on medium-band photometry. Moreover, the \ha\ emission provides
an independent measure of the stellar populations, and confirms our
findings based on just the continuum shape. This is a reassuring
result, as at the targeted redshift range \ha\ measurements are rare,
and thus our understanding of galaxies is primarily based on
photometric measurements, which use uncertain stellar population
models to interpret the nature of the continuum emission.

There are several caveats. Firstly, as indicated by the solid gray
dots and dotted gray lines, the 4000\,\AA\ breaks are corrected for
resolution (see Section 3.2 and Appendix C). As the corrections
factors are derived using the best-fit \cite{bc03} models, they may be
uncertain for galaxies which are poorly fit. Secondly, none of the
values are corrected for photometric redshift uncertainties. Thus,
depending on the exact redshift uncertainties, \dn\ and \wha\ may have
been slightly underestimated for certain SED types. For \dn\ this only
affects the SED types with strong breaks. Thirdly, \wha\ also includes
line contributions from \nt\ and \st. Their relative contribution will
be higher for metal rich systems or for cases in which an AGN is
contributing to the line emission. Thus, caution is required when
interpreting this plot.

Lastly, although \dn\ is measured over a narrow wavelength interval,
dust obscuration slightly affects the break measurement, as indicated
by the horizontal dust vector in Figure~\ref{corr}. \wha\ is also
affected by dust, as the attenuation toward the continuum emission is
generally less than the extinction towards star-forming regions
(vertical dust vector, Figure~\ref{corr}). In order to illustrate
possible offsets, the galaxies in Figure~\ref{corr} are color coded
according to their best-fit values for \av. As expected, the most
dusty galaxies, as indicated in orange, indeed have slightly stronger
4000~\AA\ breaks compared to less obscured galaxies with similar
values for \wha. However, we do not correct for this, as the values
of \av\ are poorly constrained.

\subsection{Dust and Morphologies}

Taking the measured values for $A_V$ at face value, we find that the
oldest galaxies with large \dn\ and low \wha\ typically have lower
values for \av\ than the star-forming galaxies. Interestingly, the
four youngest galaxy types with the lowest values for \dn\ and the
highest SSFR (29, 30, 31 \& 32) have similarly low extinction as
galaxies with the strongest breaks. Thus, this may imply that
star-forming galaxies are more dusty at intermediate SSFRs.

We also qualitatively investigate how the morphologies change with SED
type. When ordering the images by the strength of \dn, a few simple
observations can be made. Firstly, the morphologies of the galaxies
with the strongest 4000\,\AA\ break are rather regular. This is also
found in the local universe and at $z\sim
2.3$ \citep[e.g.,][]{vd08}. Secondly, more than half of the displayed
galaxies of the ``oldest'' SED type (1) have close neighbors,
suggestive of mergers. The sample is small though, and this
observation needs to be further assessed. Thirdly, almost all of the
displayed galaxies with the highest dust content (e.g., 15 \& 19) are
edge-on disks. SED types with similar values for \wha\ and \dn\ as the
edge-on disks, but with a lower dust content (e.g., 16 \& 18) contain
several face-on disks, which are likely the same class of objects
viewed from a different angle. When going to lower values of \dn\ and
higher values of \wha, more galaxies have irregular and clumpy
morphologies. This is consistent with what is found at
low \citep{ov08} or even higher
redshift \citep[e.g.,][]{kr09b,fo10}. A more quantitative analysis
will be presented in a future paper.

\begin{figure*}  
  \begin{center}  \includegraphics[width=0.98\textwidth]{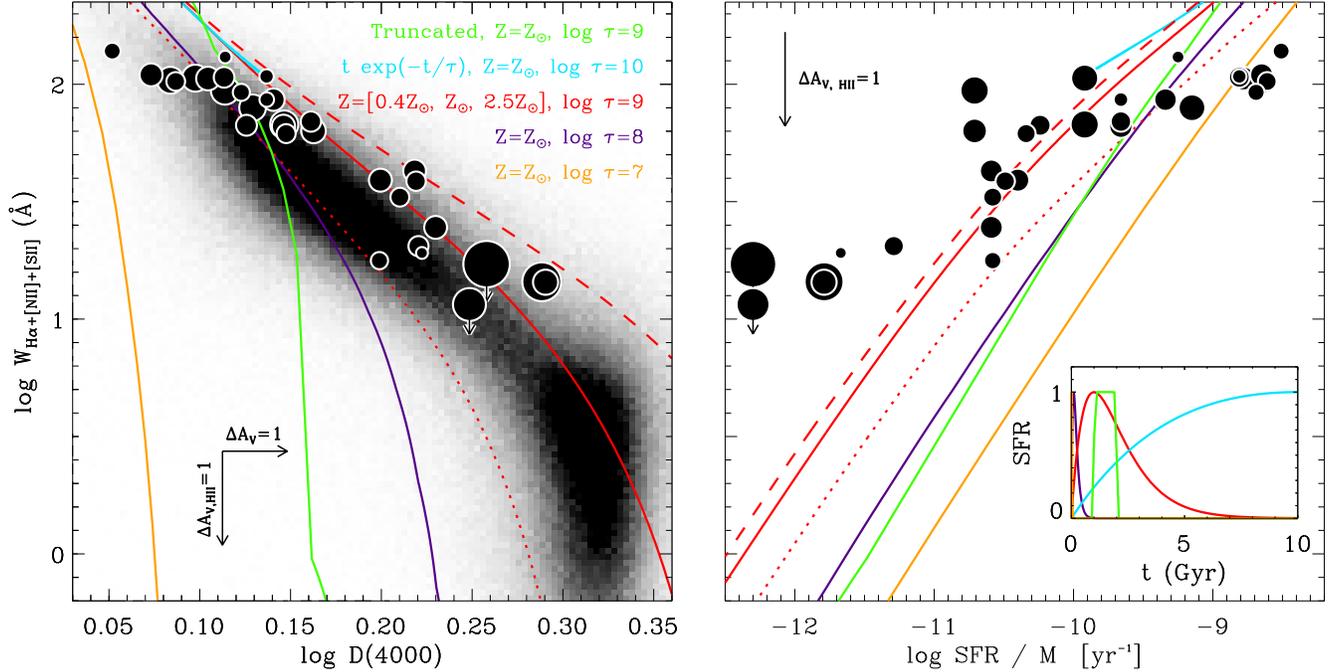}
    \caption{Left: The EW of \ha\ vs the strength of the 4000\,\AA\
      break for the 32 SED types presented in this work (solid
      circles) compared to low-redshift galaxies in the SDSS (blocked
      gray scale) and different SFHs (colored lines). We explore
      models with SFHs with gradual declining SFRs (red) and contrast
      those with SFHs with abrupt declines (yellow, green, and
      purple). The shapes of the SFHs are visualized in the inset in
      the right panel. Subsolar, solar, and supersolar models are
      represented by the dotted, solid, and dashed lines,
      respectively. The sizes of the solid circles scale with the
      number of galaxies per SED type.  For clarity, error-bars are
      not shown in this figure (see Figure~\ref{corr}). Right: The EW
      of \ha\ vs. the best-fit modeled specific SFR (assuming
      the \cite{bc03} models) for the 32 SED types in comparison to
      different SFHs. Symbols and lines are similar to the left
      panel. No similarly derived SSFRs are available for the galaxies
      in the SDSS. Models with abruptly declining SFRs predict that
      evolved galaxies at low to intermediate \dn\ and SSFRs have low
      values for \wha\ (yellow, green, and purple). Since the bulk of
      the galaxies do lie well above those curves, we conclude that
      these models do not describe the majority of the galaxies. Thus,
      the suppression of star formation at $z<2$ is generally a
      gradual process. \label{comp}}
  \end{center}  
\end{figure*}  

\section{COMPARISON TO LOW-REDSHIFT GALAXIES AND SPS MODEL PREDICTIONS}

\subsection{Comparison to the Sloan Digital Sky Survey}

In this section we compare our measurements with those of low-redshift
galaxies from the SDSS \citep{yo00}. For the SDSS galaxies we use the
spectroscopic measurements from the MPA-JHU DR7
release \citep[e.g.,][]{ka03,br04,tr04}. In Figure~\ref{comp} we show
the $0.5<z<2.0$ correlations between \wha\ and \dn\ in comparison to
the SDSS galaxies. No similarly derived SSFRs are available for the
galaxies in the SDSS and thus they are not shown in the right
panel. The sizes of the symbols in Figure~\ref{comp} reflect the
number of galaxies per composite SED.  

The relation between \wha\ and \dn\ for the galaxies  at $0.5<z<2.0$
is similar to the low-redshift SDSS galaxies. There may be a slight
offset to higher values of \wha\ for the $z\sim 1$ galaxies. This
possible offset is likely not caused by systematic errors. Photometric
redshift uncertainties would move the $0.5<z<2.0$ relation in the
opposite direction, and contamination by an [O\,{\sc ii}] emission
line would make the break at $z\sim 1$ even weaker.

We investigate possible physical explanations by comparing the
observations with different model predictions. In Figure~\ref{comp} we
show tracks for different SFHs, which are derived from the \cite{bc03}
models in combination with the relation between SFR and \ha\ as given
by \cite{ke98}. The SFHs are visualized in the inset in the right
panel in Figure~\ref{comp}. For subsolar, solar, and supersolar
metallicity we assume (N\,{\sc ii}+S\,{\sc ii})/\ha\ of 0.23, 0.8, and
0.8 respectively \cite[e.g.,][]{de02}. Comparison with model tracks
suggests that galaxies at $z\sim 1$ may have SFHs with longer
timescales or higher metallicity compared to $z\sim 0$ galaxies.

The slightly different loci of the SDSS galaxies and our composite SED
sample may also be due to systematic differences in the reddening
toward H\,{\sc ii} regions. For example, star-forming regions may be
relatively more obscured at low redshift. Higher black-hole accretion
rates at earlier times could also explain the offset location of
certain SED types at  $0.5<z<2.0$. This is further supported in the
right panel of Figure~\ref{comp}, as the location of several composite
SEDs cannot be explained by any SFH. In this context it is interesting
to note that in the SDSS many massive, quiescent galaxies are known to
host LINERs \cite[e.g.,][]{ka03b}, and thus the \wha\ of the quiescent
galaxies in SDSS likely do not reflect their star formation
activity. Follow-up spectroscopy is needed to study the nature and
origin of the line emission.

\subsection{Implication for Star Formation Histories}

It is striking that \dn\ and \wha\ correlate so well, given that \wha\
is mostly sensitive to current star formation, while \dn\ primarily
reflects the past star formation. As shown by the various model tracks
in the left panel of Figure~\ref{comp}, galaxies could in principle
occupy roughly all space in between the yellow solid and red dashed
line. However, we find that the composite SEDs and the individual SDSS
galaxies span a tight relation, close to the solid purple or dotted
red line at higher values of \wha\ and close to the red solid line for
low values of \wha. Thus, star-forming galaxies seem better explained
by short star formation timescales or lower metallicities, while the
more quiescent galaxies are better explained by SFHs with exponential
decaying times of about 1\,Gyr and solar metalliticy.

This finding suggests that the more quiescent galaxies contain stars
over a wide age range and that their star formation is gradually
declining. To further illustrate this point, we have added a truncated
model to Figure~\ref{comp}. This model has a constant SFR for about 1
Gyr, after which the star formation declines to zero. The model is
convolved by a Gaussian with a $\sigma$ of 50 Myr. Thus, the decline
occurs over roughly 0.2\,Gyr. This model predicts an abrupt decline
of \wha\ compared to the evolution of \dn, and therefore a population
of galaxies with very weak or absent \ha\ emission and weak 4000\,\AA\
breaks. However, of all composite SEDs with \dn\,$<0.24$ all have
detected \ha\ emission with a minimum value of 17.7\,\AA. 

We can further quantify this argument by considering the slow
suppression model in red and the fast suppression model in green. For
each model we derive how long it will take for the 4000~\AA\ break to
increase from log \dn\,$=0.16$ to log \dn\,$=0.24$. For the two models
this is 1.56 and 0.78\,Gyr, respectively. Thus, if only these two SFHs
would exist and both star formation timescales were equally common, we
would find that 33\% of the galaxies with log \dn\,$=[0.16,0.24]$
would have undetected \ha\ emission. However, we find that out of 778
galaxies with $0.16<$log \dn$<0.24$, none have detected \ha\ lower
than 17.7\,\AA. The non-existence of these composite SEDs implies that
SFHs with very short star-forming and quenching timescales are quite
uncommon. However, short timescales may be possible for the quiescent
galaxies that have no detected \ha\ emission (upper limits), if there
quenching occurred at $z>2$. 

We note that AGNs may complicate this argument. As we have discussed
above, the line emission, in particular for galaxies with strong
4000\,\AA\ breaks (both at $z\sim 0$ and $z\sim 1$) may originate from
actively-accreting black holes. Although it is unlikely that the line
emission in all these galaxy types is dominated by AGNs, caution is
required in the interpretation, and follow-up spectroscopic studies
are needed to uncover the origin of the ionized gas.

\section{SUMMARY AND CONCLUSIONS} \label{sec:dis}

In this paper we study the \ha\ emission line, the 4000\,\AA\ break,
and the SSFRs of $\sim$3500 galaxies at $0.5<z<2.0$, using the NMBS
high-quality medium-band photometry in the COSMOS field. The optical
and NIR medium bands are complemented by UV-to-IR broadband
photometry, leading to a total of 33 bands for each galaxy. We have
identified analogous galaxies using 22 artificial rest-frame filters
which are evenly spaced in log\,$\lambda$ between 1200 and
50\,000\,\AA, resulting in 32 galaxy subsamples. Hence, our
partitioning method is independent of stellar population models or
other assumptions. For each subsample we constructed a composite SED
by scaling and de-redshifting the observed photometry. 

The composite SEDs are of spectroscopic quality and show spectral
features, such as the (blended) \ha+\nt+\st\ and H$\beta$+[O\,{\sc
iii}] emission lines, the Balmer or 4000\,\AA\ breaks, Mg{\sc\,ii}
absorption at 2800~\AA, the continuum break at 2640~\AA, and the dust
absorption feature at 2175\,\AA. Most of these features are not
detected in the SEDs of individual galaxies. Thus, the increased S/N
and sampling of the composite SEDs enable the detection and
measurement of spectral features, which would normally require
spectroscopic data. Therefore these SEDs can be used to study the
stellar populations and emission line characteristics of typical
galaxies in a very detailed way, which is not possible when the
photometry of individual galaxies is considered
separately. Additionally, the shape and linewidth of the \ha\ emission
line yield an indepedent measure of the photometric redshift
errors. Consistent with comparison to spectroscopic redshifts, we find
an accuracy of $\Delta z \lesssim 0.02 \times (1+z)$.

In the present work, we measure \wha\ and \dn\ for all composite SEDs
and compare the full SED shapes with SPS models. Most SED types are
equally well fit by the \cite{ma05} and \cite{bc03} SPS models. The
exception, however, are post-starburst galaxies. As has previously
been shown in \cite{kr10}, the \cite{ma05} models cannot reproduce the
full SED shape of post-starburst galaxies simultaneously. They
overpredict the NIR luminosity, implying that these models give too
much weight to TP-AGB stars.

We find that \wha, \dn\ and the best-fit SSFR are all strongly
correlated, such that galaxies with shallow breaks have high values
for \wha\ and are best fit by stellar population models with high
SSFRs. On the other hand, galaxies with strong 4000~\AA\ breaks have
low values for \wha\ and are best fit by SPS models with low SSFRs.
This is a remarkable result given that the relations are derived
without spectroscopy, and solely based on medium-band
photometry. Moreover, the \ha\ lines provide an independent measure of
the stellar populations and confirm the results based on just the
continuum shape. \ha\ measurements at the targeted redshift range are
very challenging with spectroscopy, as the line is shifted to NIR
wavelengths. At this moment, spectroscopic \ha\ measurements beyond
$z=0.5$ are only availably for small or biased galaxy
samples \citep[e.g.,][]{sh04,er06c,kr08b,fo09}, and consquently our
understanding of galaxies at this epoch is primarily based on
photometric measurements, which use uncertain stellar population
models to interpret the nature of the continuum emission. The
composite SEDs -- for the first time -- open up the possibility to
efficiently study \ha\ emission for large, magnitude-limited galaxy
samples, providing a crucial assessment of our photometric studies.

We also study the rest-frame optical morphologies of a subsample of
galaxies around $z\sim1$ for each SED type. Interestingly, we find
that with increasing \dn, galaxies appear to change from irregular and
clumpy, to fairly regular disks, to regular early-type systems at the
highest \dn.  Strikingly, the most dusty galaxies are typically
edge-on disks.

The relation between \wha\ and \dn\ at $0.5<z<2.0$  is similar to the
low-redshift relation, as determined from the SDSS galaxy
sample. There may be a slight offset, such that at fixed value
for \dn, the \wha\ of the  $0.5<z<2.0$ galaxies is slightly higher
than for the low-redshift galaxies. We use simple models to interpret
this difference, and show that the rate at which the SFR is declining
can explain the difference in \wha\ at fixed \dn; specifically, the
SFRs in $z\sim 1$ galaxies may decline more gradually than at low
redshift. The elevated \wha\ may also be explained by higher black
hole accretion rates or higher metallicities at earlier times, or by
relatively more obscured H\,{\sc ii} regions at lower redshift.

It is remarkable that \wha\ and \dn\ are so tightly correlated, given
that \wha\ is mostly sensitive to current star formation and \dn\ is a
measure of the older stellar population, and thus the past star
formation. SPS model predictions show that SFHs that decline rapidly
after their major burst -- such as truncated SFHs or SFHs with short
timescales -- predict a steeper decline of \wha\ compared to \dn\ than
what is observed for the galaxies with intermediate to low values
of \wha. The combination of \dn\ and \wha\ and the best-fit star
formation timescales both indicate long star formation timescales of
$\sim1$\,Gyr (for a SFH parametrized by $t$\,exp($-t/\tau$)) for older
galaxy types. This suggests that the suppression of star formation at
$z<2$ is generally not an abrupt, but rather a gradual process. Our
finding is consistent with the predictions by high-resolution smoothed
particle hydrodynamics (SPH) simulations of massive elliptical
galaxies starting from $\Lambda$CDM initial conditions \citep{na07}.

The SEDs that we presented here can be used for many additional
purposes. For example, we can measure the relative strength of
emission lines, the strength of absorption features (e.g., Mg\,{\sc
ii} at 2800\,\AA), and constrain the dust extinction curve including
the 2175\,\AA\ dust absorption feature. This paper briefly discussed
the morphologies of the different SED types. A more quantitative
analysis, with measurements of sizes and ellipticities will be
presented in a future paper. For example, by comparing the sizes of
galaxies in several evolutionary phases, we can assess theories
proposed to explain the size growth of elliptical
galaxies \citep[e.g.,][]{vd08,na07,na09,be09,ho09,ho10}. Comparison
with other wavelength regimes also offers many interesting
possibilities. For instance, X-ray stacks allow the study of AGN
demographics among the different types.

Due to the low resolution of the NMBS, it is unfortunately not
possible to deblend the emission lines H$\alpha$, [N\,{\sc ii}],
and \st\ or H$\beta$ and [O\,{\sc iii}]. Thus, NIR spectroscopy is
needed to test our H$\alpha$ measurements, to derive emission line
ratios, and to study the contribution from AGNs to the emission line
fluxes. Such measurements would also be very valuable to study the
metallicities and Balmer decrements within and among the different
samples. Altogether, our composite SEDs in combination with follow-up
spectroscopy and other wavelength data enable many different studies
to better understand the origin and evolution of galaxies.

\acknowledgements We thank the members of the NMBS team for their help 
with the observations and construction of the catalogs, Charlie
Conroy, Ryan Foley, and Jeremiah Ostriker for useful discussions, and
the COSMOS and AEGIS teams for the release of high-quality
multi-wavelength data sets to the community. M. K.  acknowledges
support of a Clay Fellowship administered by the Smithsonian
Astrophysical Observatory.

\appendix

\section{APPENDIX A: COMPLETENESS}

\begin{figure*} 
  \begin{center} 
    \includegraphics[width=0.98\textwidth]{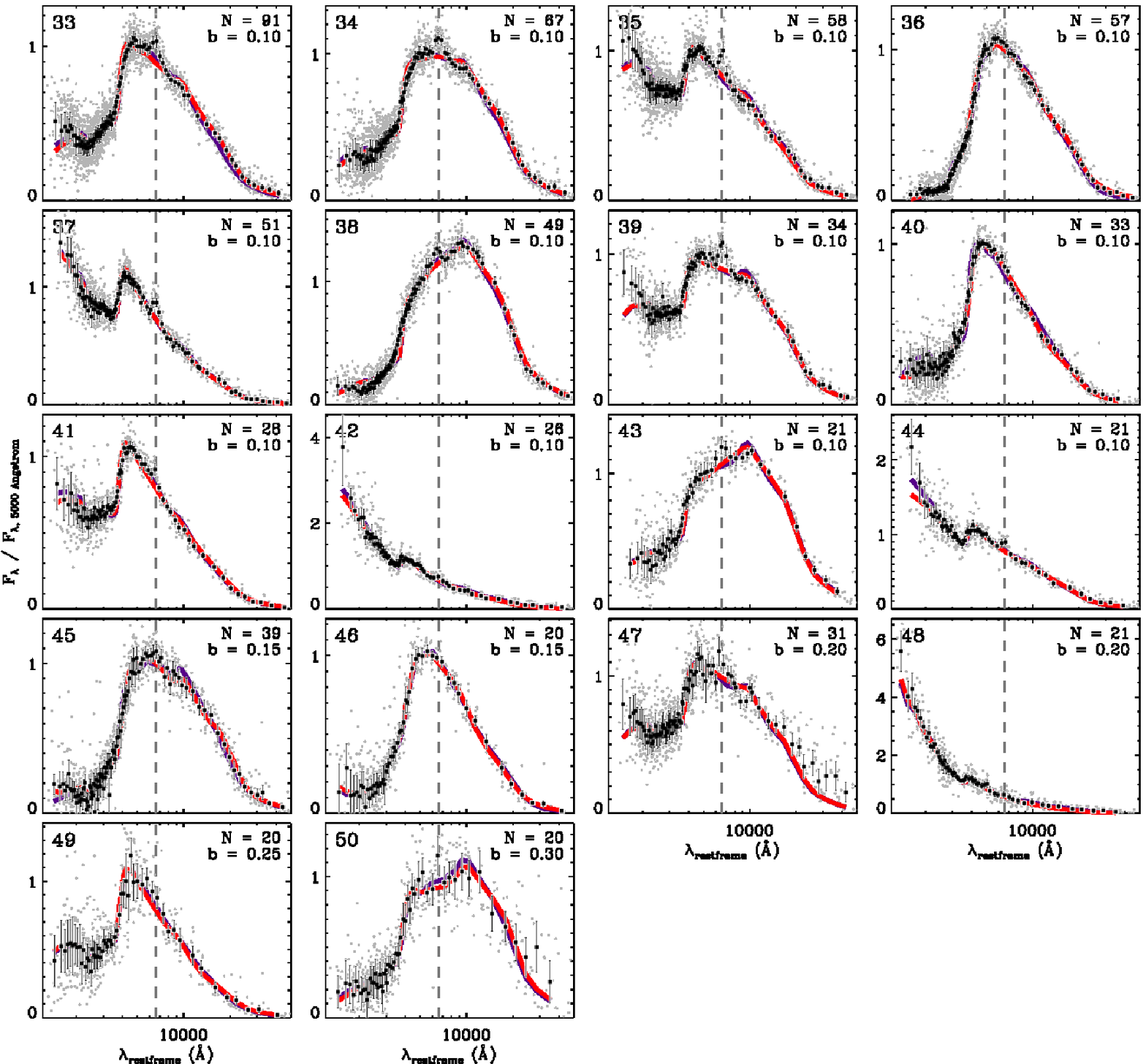}

    \caption{Composite SEDs of the remaining 17\% of the parent
      sample, that were not included in the original 32 galaxy
      composite SEDs in Figure~\ref{compilation}. The symbols and
      lines are the same as in Figure~\ref{compilation}. The remaining
      galaxies are binned by increasing the value of $b$ in Equation
      (1), until less than 20 galaxies are left. Most SED types look
      similar to ones in the initial samples, though with more
      scatter. However, there are several types which are different,
      such as SED 42, 44 and 48. Thus, the most un-obscured
      star-forming galaxies are likely missed in our primary
      sample. \label{rem} } 
  \end{center}
\end{figure*}

\begin{figure*} 
  \begin{center} 
    \includegraphics[width=0.65\textwidth]{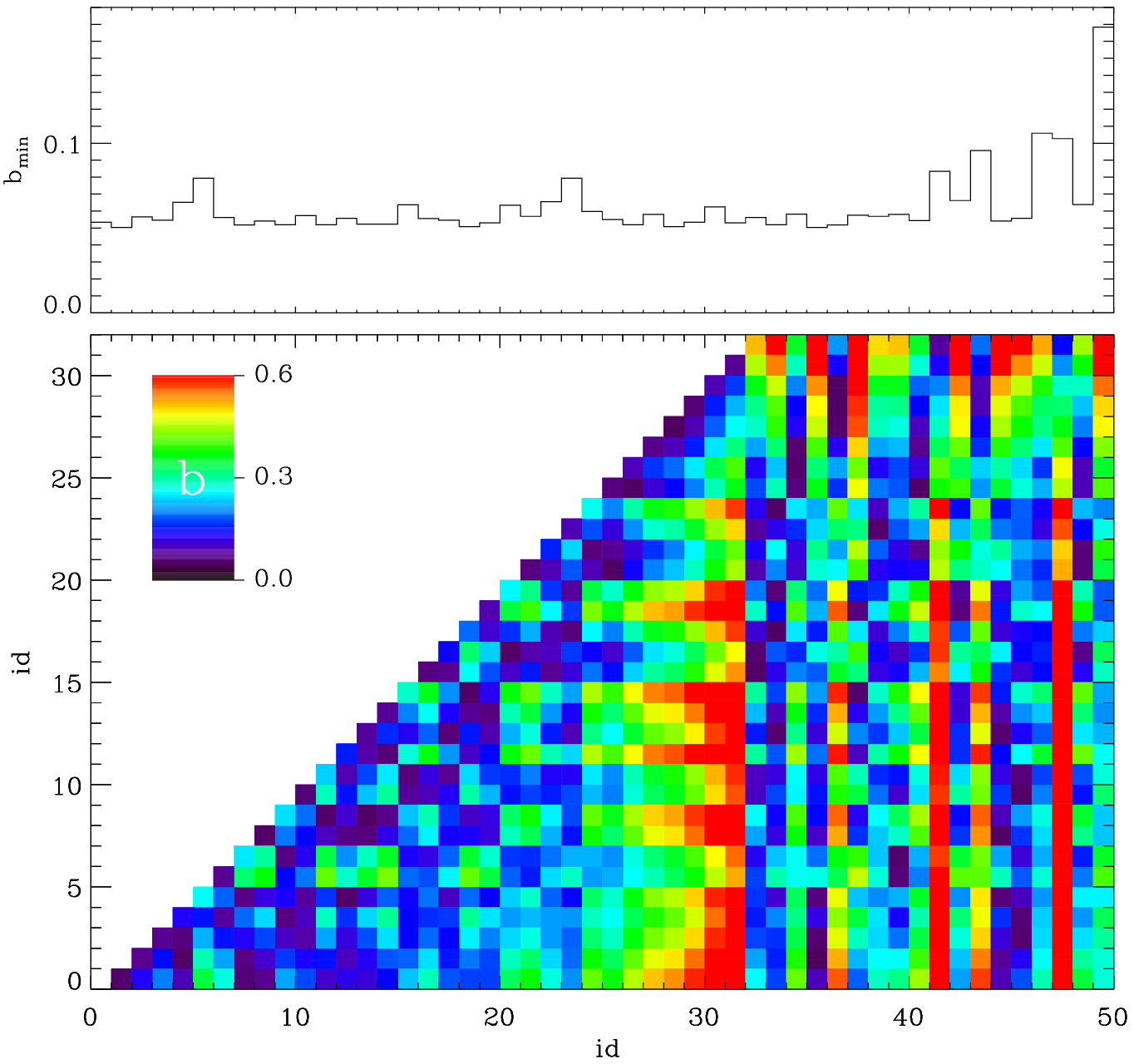} 

    \caption{Bottom: The value of the correlation quantity $b$ (see
      Equation 1) between the primary galaxies of the different
      subsamples. Top: the minimum value of $b$ for the primary galaxy
      of a certain SED types, when compared to all other primaries in
      the main sample (first 32 SEDs). \label{b}} 

  \end{center}
\end{figure*}

The composite SED collection of 32 template constitutes 83\% of the
$\sim$4200 galaxies in our parent sample. Here, we assess the
completeness and discuss which SED types we are possibly missing. For
the 17\% of the remaining galaxies we use the same technique to find
analogous galaxies, but now taking $b=0.1$ in Equation (1). We divide
up the sample, such that each primary has at least 19 analog galaxies
and construct 12 composite SEDs with this value of $b$. They are
presented in Figure~\ref{rem} (33-44). Next we increase $b$ to 0.15,
0.20, 0.25 and 0.30 until the remaining sample is less than 20
galaxies. The six composite SEDs constructed with values of $b>0.1$
are shown in Figure~\ref{rem} as well. 

Many SED types in Figure~\ref{rem} look similar to one of the 32
galaxy types in the main sample. However, SEDs 42, 44 and 48 are not
represented in the 32 galaxy types. Thus, the most un-obscured
star-forming galaxies are likely missed in our primary sample. Their
SSFRs are in the order of log SSFR = [-7,-8]. However, as their Balmer
breaks are very shallow and as their Lyman breaks are not in the
observed window for $0.5<z<2.0$,  their photometric redshifts have
large uncertainties. Thus, the measured \wha\ for these galaxies is
highly uncertain. Galaxy types with high values for $b$ give less
insight into the remaining galaxy population, as they are likely
composed of intrinsically different SED types. 

To do a more quantitative analysis, we relate all primary galaxies
with the primaries of the main sample in Figure~\ref{b}. In the bottom
panel we indicate the values for $b$ (see Equation 1) between the
different SED types. In the top panel we show the minimum value of $b$
for each SED type, when compared to all other primary galaxies of the
main sample. This figure illustrates that our main sample is well
sampled, with typical $b$ values of $\sim$0.05-0.06 between the
closest types. SED types 6 and 24 have slightly higher values with
$b\sim0.08$. The SED types in the $b=0.1$ sample (SED type 33-44,
Figure~\ref{rem}) relate to the main sample with values of
$b\sim0.06$. The exceptions are SED types 42 and 44. As mentioned
above, these galaxies are indeed not represented in our main sample.
The SED types with higher values of $b$ are less correlated to the
primaries of the main samples. For some types (e.g., 48) this means
that their SED is very different from the main sample, while for other
types this may reflect the fact that the large variation of galaxies
within one type leads to unphysical composite SEDs.

\section{APPENDIX b: STACKING TEST}

\begin{figure} 
  \begin{center}     
    \includegraphics[width=0.31\textwidth]{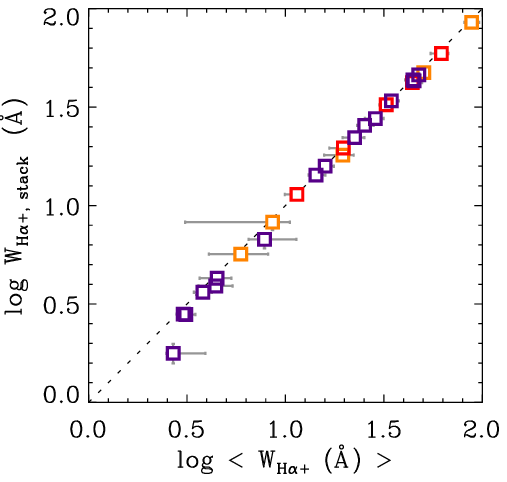} \hspace{0.02in} 
    \includegraphics[width=0.31\textwidth]{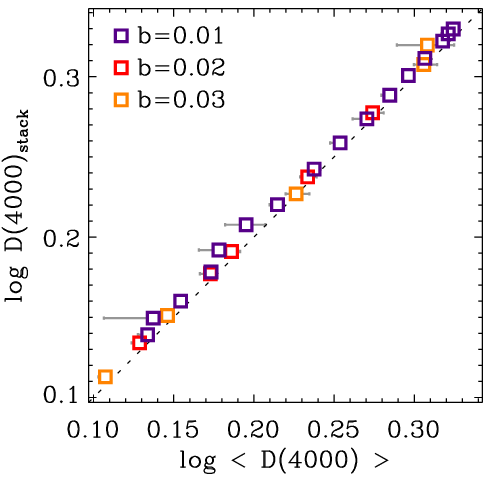}  \hspace{0.02in} 
    \includegraphics[width=0.31\textwidth]{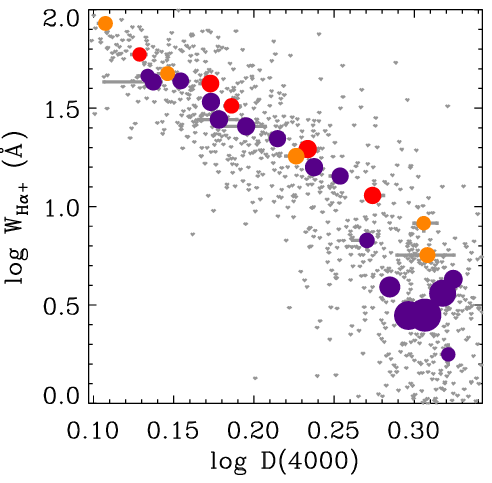}   
  \end{center}   

  \caption{Comparison of the average values of individual galaxies
    with the value as measured from the stacked spectrum of the same
    individual galaxies, for both \wha\ (left) and \dn\ (middle). For
    this test we use $\sim$1000 galaxies in SDSS at $z\sim0.1$, and
    construct the stacked spectra in the same way as for the higher
    redshift sample. The right panel shows \wha\ vs. \dn, with the
    individual measurements represented by the small gray dots and the
    stacked values represented by the solid circles. The sizes of the
    solid circles scale with the number of galaxies in the stack. The
    colors of the symbols indicate the value for $b$ (see Equation 1)
    of the stacked spectrum. This figure illustrates that with our
    stacking method we recover the average properties of the
    individual galaxies. \label{stacktest}}   
    
\end{figure}   

In this appendix we assess whether our stacking technique indeed
yields the average properties of the individual galaxies in the
subsamples. We do so by applying our stacking technique to galaxy
spectra in the SDSS. We start with 10\,000 galaxy spectra in the SDSS
and select all galaxies at $0.095<z<1.05$. For each galaxy we
determine the fluxes in the rest-frame filters (shown in
Figure~\ref{filters}), which overlap with the spectra. Next, we
correlate all galaxies with each other using Equation 1. As the
galaxies are matched using only three filters, we start with a smaller
value for $b$ of 0.01 to find analogs. Similar as for the higher
redshift sample, we increase the value for $b$, requiring that each
SED type consists of a minimum of 20 galaxies. We increase $b$ until
nearly all galaxies are part of an SED bin. In total we have 26
subsamples for which we construct the composite SEDs.

We measure \dn\ in the composite spectra and compare these to the mean
of the individual measurements in Figure~\ref{stacktest}. The
individual measurements are adopted from the MPA-JHU DR7 catalogs. The
color of the datapoints reflect the value of $b$, and thus the scatter
of the properties within the subsamples. The errorbars on the mean
properties of the individual galaxies are derived using
bootstrapping. We also calculate the errors on the stacked values
using simulated spectra, which are constructed using the noise
spectra.

There is good agreement between the break measurements for
\dn\ for the stack and the individual systems, with a small
systematic offset toward higher values for the stack. The different
measurements for \wha\ are also in very good agreement, except for at
the lowest value of \wha\ of 2-3~\AA. This is reassuring, in
particular given the fact that both measurements use different
techniques to correct for the underlying Balmer absorption. The
agreement does not worsen for larger values of $b$; even for composite
spectra which are constructed from galaxies which exhibit a large
intrinsic scatter in SEDs, the stacks still recover the average of the
individual measurements.

\section{APPENDIX c: \dn\ CORRECTION}

\begin{figure}    
  \begin{center}     
    \includegraphics[width=0.44\textwidth]{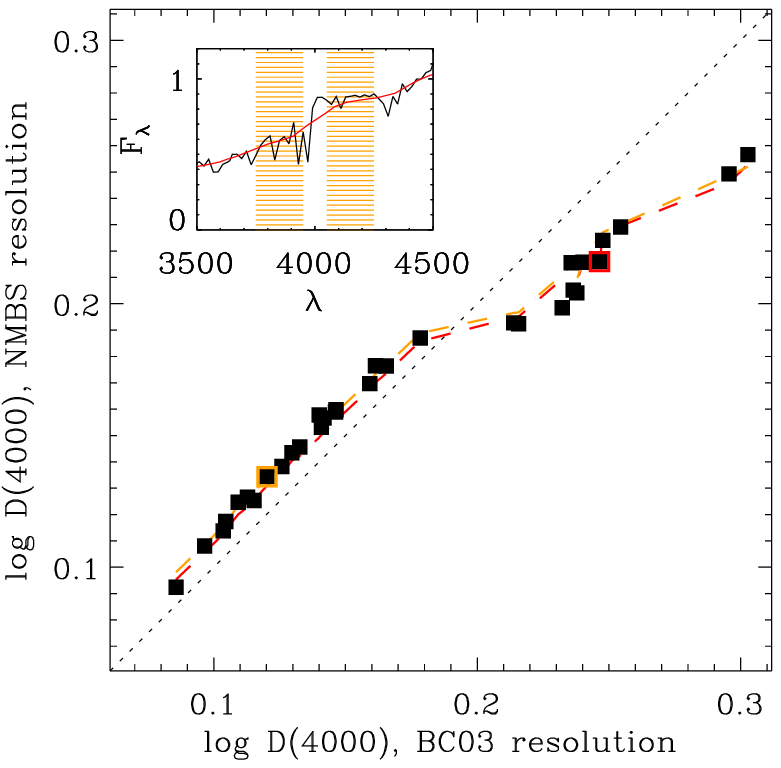} \hspace{0.4in}   
    \includegraphics[width=0.44\textwidth]{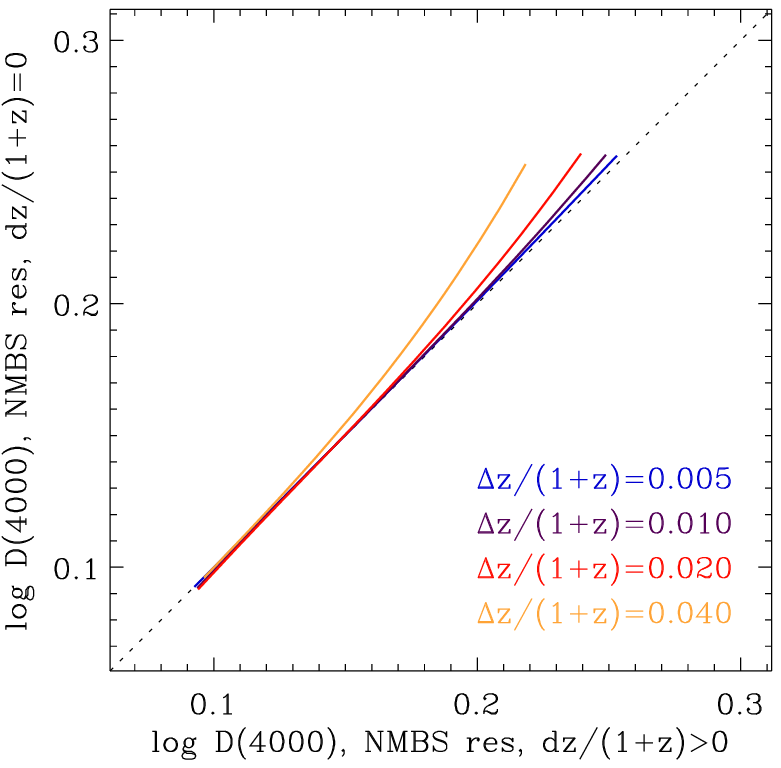}    
  \end{center}  

  \caption{Left: Measurements for \dn\ derived from the
    best-fit \cite{bc03} models to the composite SED at the original
    resolution versus \dn\ measured after convolving the best-fit
    model with the composite filter curves.  For each composite SED we
    derive a resolution correction factor based on the ratio of these
    two \dn\ measurements. The correction factor depends on the
    strength of the break and the composite filter curves. The dashed
    lines represent the correlations for two particular filter
    sets. Right: The effect of photometric redshift errors on the
    break measurement. We do not correct for this effect, as the exact
    redshift uncertainties for each SED type are unknown. Nonetheless,
    with this exercise we caution that the measurements for \dn\ for
    larger breaks may be underestimated.\label{dncorr}}  

\end{figure}

The composite SEDs allow the direct measurement of the 4000\,\AA\
break. In this paper we use the index \dn\ by \cite{br83} to quantify
its strength. This index measures the difference in flux in small
wavelength regions bracketing the break, as indicated by the orange
shaded areas in the inset in Figure~\ref{dncorr}. However, this
measurement will by affected by the low resolution of the composite
SED. The inset in Figure~\ref{dncorr} shows a synthetic SED in the
region around the break. The red curve shows this same SED, but now
convolved by the composite filter curves. This figure clearly
illustrates that for this case the convolved SED will yield a slightly
lower value for \dn\ than the actual SED.

We can correct for this effect as we know the composite filter curves
for all SEDs. For each SED type we derive a correction factor using
its best-fit stellar population model. First, we measure the break in
the original best fit, and then we derive \dn\ for the same model, but
now convolved with the composite filter curves. We show both these
measurements for each SED type in Figure~\ref{dncorr}. We obtain the
correction factor by dividing them by each other, and we assume a 5\%
error on this factor.

The correction factor is dependent on the actual value of \dn\ and the
sampling of the composite filter curves. For example,
Figure~\ref{dncorr} shows that for small breaks the correction factors
are close to unity. This is not surprising, as for these galaxies \dn\
measures the slope of the continuum, and this is less affected by
resolution. However, the larger the break, the larger the correction
factor. The dependence on the filter curves is illustrated using the
dashed colored lines. These lines indicate the correction for
different values of \dn\ for a particular composite filter set. For
SEDs with less composite filters the corrections are larger. Note that
the break measurement is not always consistent with that of the
best-fit model, and thus the distribution of \dn\ in
Figure~\ref{dncorr} is slightly different from the measured
distribution. 

The break measurements are also affected by photometric redshift
uncertainties. In the right panel in Figure~\ref{dncorr} we show the
effects on the break measurements for different values of $\Delta
z/(1+z)$. The widths of the blended \ha\ emission lines (see
Figure~\ref{hadz}) indicate a photometric redshift uncertainty of
$\Delta z/(1+z)<0.02$. Comparison with spectroscopic redshifts shows
that the photometric redshifts in the NMBS have an uncertainty of
$\Delta z/(1+z)=0.01-0.02$ \citep{wh11}. However, the uncertainties
will depend on SED type \citep[e.g.,][]{kr08a}, and as the exact
scatter per SED type is unknown, we have decided not to correct the
break measurements for photometric redshift uncertainties. This means
that the values for \dn, in particular for galaxies with strong breaks
may be underestimated.

\end{document}